\documentclass[]{interact}
\usepackage[nolists,tablesfirst]{endfloat}
\usepackage{bm}
\allowdisplaybreaks
\usepackage{slashed}
\usepackage{float}
\usepackage{nicefrac}
\usepackage{dcolumn}
\newcolumntype{.}{D{x}{}{-1}}
\newcolumntype{w}[1]{D{.}{.}{#1}}
\usepackage{booktabs,tabularx}
\newcommand\mc[1]{\multicolumn{1}{c}{#1}}
\newcommand{\phiel}{\phi_\mathrm{el}}
\newcommand{\psiS}{\phi_{\Sigma^{\!+}}\!}
\newcommand{\Eel}{\mathcal{E}_{\mathrm{el}}}
\newcommand{\Hel}{H_{\rm el}}
\newcommand{\nr}{\vec{\nabla}_{\!R}}
\newcommand{\bk}[2]{\bigl\langle#1\big|#2\bigr\rangle}
\begin{document}
\articletype{Research Article}

\title{Nonadiabatic corrections to electric quadrupole transition rates in H$_2$}

\author{
\name{Krzysztof Pachucki\textsuperscript{a} and Michał Siłkowski\textsuperscript{b}}
\affil{\textsuperscript{a} Faculty of Physics, University of Warsaw, Pasteura 5, 02-093 Warsaw, Poland \\
\textsuperscript{b}\thanks{CONTACT M. Si{\l}kowski. Email: michal.silkowski@amu.edu.pl} Faculty of Chemistry, Adam Mickiewicz University in Pozna{\'n}, Uniwersytetu~Poznańskiego~8, 61-614 Pozna{\'n}, Poland}
}

\date{\today}

\maketitle

\begin{abstract}
We derive formulas and perform calculations of nonadiabatic corrections to rates of electric quadrupole transitions in the hydrogen molecule.
These corrections can be represented in terms of a single curve $D^{(1)}(R)$, similarly to the Born-Oppenheimer one, $D^{(0)}(R)$,
derived originally by Wolniewicz. Numerical results change E2 transition rates for the fundamental band by as much as 0.4 - 12\% depending on rotational quantum numbers.
\end{abstract}


\section{Introduction}

Most of the calculations of molecular properties are performed in the Born-Oppenheimer (BO) approximation. When higher precision is needed, e.g., for accurately measured transition energies in H$_2$ \cite{Holsch:19, Beyer:19,Cozijn:23, Stankiewicz:25}, one includes adiabatic and nonadiabatic corrections. These corrections can be calculated systematically within the so-called nonadiabatic perturbation theory (NAPT)~\cite{nonad1,nonad2}. Using NAPT, one can calculate not only corrections to rovibrational energies, but also to many other physical properties \cite{magmol},
like the nuclear magnetic shielding. This allowed for the most accurate determination of deuteron and triton magnetic moments~\cite{triton2015}.

In this work, we derive a formula for the leading nonadiabatic correction to rates of electric quadrupole transitions in H$_2$.
This nonadiabatic correction can be represented in terms of a single function $D^{(1)}(R),$ which can be added to the BO quadrupole moment function $D^{(0)}(R)$, calculated by Wolniewicz in Ref.~\cite{Wolniewicz98}, recalculated by Komasa \cite{komasam1} using ECG functions, and significantly improved here using an enhanced version of H2SOLV code \cite{h2solv}, which employs the Kołos-Wolniewicz (KW) functions.
Apart from many applications in astrophysics, these accurate rates can be used in the primary thermometry for temperatures as low as 10 K \cite{Stankiewicz:25, shuiminghu24, shuiminghu25,Lisak:25}. In particular Authors of Ref. \cite{shuiminghu25} propose to  measure the ratio of two transition rates from the same vibrational band 
to achieve higher accuracy for temperature due to cancellation of experimental and theoretical uncertainties.  
Hydrogen molecule is the best for this purpose, because transition rates can be calculated very accurately from first principles, 
and this work is the first step toward this accurate termometry.

\section{Nonadiabatic perturbation theory}
Following Ref. \cite{magmol}, let us start from a nonrelativistic Hamiltonian for a neutral diatomic molecule
\begin{equation}
H = \sum_a\frac{\vec p_a^{\;2}}{2\,m} +\frac{\vec p_A^{\;2}}{2\,m_A}
+\frac{\vec p_B^{\;2}}{2\,m_B} + V, 
\label{01}
\end{equation}
where the summation index $a$ enumerates all electrons, and $A$ and $B$ refer to nuclei,  and atomic units are used throughout the paper. 
To derive formulae for nonadiabatic effects, one must fix the reference frame. We start with the laboratory frame $\{\vec r_A,\,\vec r_B,\,\vec r_a\}$, and subsequently change variables to $\{\vec R,\,\vec R_G,\,\vec x_a\}$
according to
\begin{align}
\vec r_A =&\ \vec R_G+\epsilon_B\,\vec R,\label{02}\\
\vec r_B =&\ \vec R_G-\epsilon_A\,\vec R,\label{03}\\ 
\vec r_a =&\ \vec R_G+\vec x_a, \label{04}
\end{align}
with  the relative position of nuclei
$\vec R = \vec r_A-\vec r_B$, and the origin of the new frame, 
$\vec R_G = \epsilon_A\,\vec r_A + \epsilon_B\,\vec r_B$, where
$\epsilon_A+\epsilon_B=1$, chosen anywhere on the internuclear axis.
The conjugate momenta are related by
\begin{align}
{\vec p}_A =&\ \epsilon_A\,\vec P_G + {\vec P} -\epsilon_A\,\sum_a\vec q_a,\label{05}\\
{\vec p}_B =&\ \epsilon_B\,\vec P_G - {\vec P} -\epsilon_B\,\sum_a\vec q_a,\label{06}\\
{\vec p}_a =&\ \vec q_a, \label{07}
\end{align}
where $\vec P = -i\,\vec\nabla_R$ and $\vec q_a = -i\,\vec\nabla_{x_a}$.
The nonrelativistic wave function with vanishing total momentum does not depend on $\vec R_G$, so $\phi = \phi(\vec x_a,\,\vec R)$. 
We now choose the center of the reference frame at the nuclear mass center,
\begin{align}
      \epsilon_A =&\ \frac{m_A}{m_A+m_B}, & \quad \epsilon_B =&\ \frac{m_B}{m_A+m_B}\,,
\end{align}
and split the Hamiltonian into the electronic and nuclear parts,
\begin{align}
H =&\ H_{\rm el} + H_{\rm n},\label{10}\\
H_{\rm el} =&\ \sum_a \frac{\vec q_a^{\;2}}{2\,m} + V, \label{11}\\
H_{\rm n} =&\ \biggl(\frac{1}{2\,m_A}+\frac{1}{2\,m_B}\biggr)\,\vec P^{\;2}
+\frac{1}{2\,(m_A+m_B)}\,
\Bigl(\sum_a\vec q_a\Bigr)^2 =
H_{\rm n}' + H_{\rm n}'' .\label{12}
\end{align}
This form of the nuclear Hamiltonian is convenient for the
calculation of nonadiabatic effects. 

Next, the total angular momentum operator $\vec J$ does not depend on the reference point, because we assumed the vanishing total 
momentum for molecular states. Therefore we chose for convenience $\vec R_G$ as a reference point and $J$ becomes
\begin{equation}
\vec J = \sum_a (\vec r_a-\vec R_G)\times\vec p_a + 
         (\vec r_A -\vec R_G)\times\vec p_A +
         (\vec r_B -\vec R_G)\times\vec p_B\,. \label{13}
\end{equation}
In new variables 
\begin{equation}
\vec J = \sum_a \vec x_a\times\vec q_a +\vec R\times\vec P 
\equiv \vec J_{\rm el}+\vec J_{\rm n},
\label{14}
\end{equation}
$J$ can also be split into electronic $\vec J_{\rm el}$ and nuclear $\vec J_{\rm n}$ parts.

Let us now introduce NAPT \cite{nonad1,nonad2}.
The total nonrelativistic wave function $\phi$ of an arbitrary molecule
is the solution of the stationary Schr\"odinger equation 
\begin{equation}
[H-E]\,|\phi\rangle = 0\,, \label{15}
\end{equation}
with the Hamiltonian $H$ being a sum of the
electronic $H_{\rm el}$ and nuclear $H_{\rm n}$ parts, Eq.\,\eqref{12}.
In the adiabatic approximation the wave function $\phi=\phi_{\rm a}$, where
\begin{equation}
\phi_{\rm a}(\vec x,\vec R) = \phi_{\rm el}(\vec x) \; \chi(\vec R), \label{16}
\end{equation}
is represented as a product of the electronic wave function $\phi_{\rm el}$
and the nuclear wave function~$\chi$. We note that $\phi_{\rm el}$
depends parametrically on the nuclear relative coordinate $\vec R$.
The electronic wave function obeys the clamped nuclei electronic 
Schr\"odinger equation 
\begin{equation}
\bigl[H_{\rm el}-{\cal E}_{\rm el}(R)\bigr]\,|\phi_{\rm el}\rangle = 0, \label{17}
\end{equation} 
while the nuclear wave function is a solution to the Schr\"odinger equation 
in the effective potential generated by electrons
\begin{equation}
\bigl[ H_{\rm n} +{\cal E}_{\rm a}(R)+{\cal E}_{\rm el}(R)-E_{\rm a}\bigr]\,
|\chi\rangle = 0\,, \label{18}
\end{equation} 
where the adiabatic correction ${\cal E}_{\rm a}(R)$  is
\begin{equation}
{\cal E}_{\rm a}(R) = \bigl\langle\phiel|H_{\rm n}|\phiel\bigr\rangle_{\rm el}\,.
\label{19}
\end{equation}
In NAPT, the total wave function is the sum of the adiabatic solution and a nonadiabatic correction
\begin{equation}
\phi = \phi_{\rm a} + \delta\phi_{\rm na} = \phi_{\rm el}\,\chi + \delta\phi_{\rm na}\,.
\label{20}
\end{equation}

The nonadiabatic correction $\delta\phi_{\rm na}$ is  decomposed into two parts,
\begin{equation}
\delta\phi_{\rm na} = \phi_{\rm el}\,\delta\chi + \delta'\phi_{\rm na}\,,
\label{21}
\end{equation}
obeying the following orthogonality conditions
\begin{align}
\langle\delta'\phi_{\rm na}|\phi_{\rm el}\rangle_{\rm el} =&\ 0\,,\label{22}\\
\langle\delta\chi|\chi\rangle =&\ 0\,,\label{23} 
\end{align}
which imply the normalization condition $\langle\phi_a|\phi\rangle=1$.
In the first order of NAPT one has
\begin{align}
|\delta'\phi_{\rm na}\rangle^{(1)} =&\ 
\frac{1}{({\cal E}_{\rm el}-H_{\rm el})'}\,
H_{\rm n}\,|\phi_{\rm el}\,\chi\rangle,
\label{24}
\end{align}
where $1/({\cal E}_{\rm el}-H_{\rm el})'$ denotes the resolvent with
the reference state $\phi_{\rm el}$ subtracted out. This correction to the wave function
can be used to derive the finite nuclear mass corrections to various matrix elements.
Consider the Hermitian electronic operator $Q$ 
(without derivatives with respect to nuclear variables) and its matrix element between  (different) rovibrational states.
In the BO approximation, this matrix element can be represented 
in terms of the diagonal electronic matrix element nested in the nuclear matrix element, namely
\begin{align}
\langle Q\rangle^{(0)} \equiv&\
\langle\phi_{\rm el}\,\chi_{\rm f}|Q|\phi_{\rm el}\,\chi_{\rm i}\rangle 
= \langle\chi_{\rm f}|\langle Q\rangle_{\rm el}^{(0)}|\chi_{\rm i}\rangle, \label{25}
\end{align}
where
\begin{align}
\langle Q\rangle_{\rm el}^{(0)} \equiv&\ \langle Q\rangle_{\rm el}  
= \langle\phi_{\rm el}|Q|\phi_{\rm el}\rangle.
\label{26}
\end{align}
We will show  that the same holds for the leading nonadiabatic corrections, which is
\begin{align}
\langle Q\rangle^{(1)} =&\ 
\langle\phi_{\rm el}\,\chi_{\rm f}|H_{\rm n}\,
\frac{1}{({\cal E}_{\rm el}-H_{\rm el})'}\,
Q|\phi_{\rm el}\,\chi_{\rm i}\rangle 
+\langle\phi_{\rm el}\,\chi_{\rm f}|Q\,
\frac{1}{({\cal E}_{\rm el}-H_{\rm el})'}\,
H_{\rm n}|\phi_{\rm el}\,\chi_{\rm i}\rangle.
\label{27}
\end{align}
Namely,
\begin{align}
\langle Q\rangle^{(1)} =&\ 
\int d^3R\,\biggl\{(\chi^*_{\rm f}\,\chi_{\rm i})\,\biggl[
\langle H_{\rm n}\,\phi_{\rm el}|
\frac{1}{({\cal E}_{\rm el}-H_{\rm el})'}\,Q\,|\phi_{\rm el}\rangle
+\langle\phi_{\rm el}|Q\,\frac{1}{({\cal E}_{\rm el}-H_{\rm el})'}\,
|H_{\rm n}\,\phi_{\rm el}\rangle\biggr]
\nonumber \\ &\ 
-\frac{
\vec\nabla\bigl(\chi^*_{\rm f}\,\chi_{\rm i}\bigr)}{2\,m_{\rm n}}\,
\biggl[
\langle\vec\nabla_R\phi_{\rm el}|
\frac{1}{({\cal E}_{\rm el}-H_{\rm el})'}\,Q\,|\phi_{\rm el}\rangle
+\langle\phi_{\rm el}|Q\,\frac{1}{({\cal E}_{\rm el}-H_{\rm el})'}\,
|\vec\nabla_R\phi_{\rm el}\rangle\biggr]
\label{28}\\ &\ 
-\frac{
\Bigl(\chi_{\rm i}\,\vec\nabla\chi^*_{\rm f}-
\chi_{\rm f}^*\,\vec\nabla\chi_{\rm i}\Bigr)}{2\,m_{\rm n}}\,
\biggl[
\langle\vec\nabla_R\phi_{\rm el}|
\frac{1}{({\cal E}_{\rm el}-H_{\rm el})'}\,Q\,|\phi_{\rm el}\rangle
-\langle\phi_{\rm el}|Q\,\frac{1}{({\cal E}_{\rm el}-H_{\rm el})'}\,
|\vec\nabla_R\phi_{\rm el}\rangle\biggr]\biggr\}\nonumber.
\end{align}
Assuming $Q$ is invariant with respect to time reversal, the third term vanishes, and with the help of integration by parts, we obtain
\begin{align}
\langle Q \rangle^{(1)} =&\ \langle\chi_{\rm f}|
           \langle Q \rangle^{(1)}_{\rm el} |\chi_{\rm i}\rangle, \label{29}\\
\langle Q \rangle^{(1)}_{\rm el} =&\ \langle\phi_{\rm el}|
\stackrel{\leftrightarrow}{H_{\rm n}}
\frac{1}{({\cal E}_{\rm el}-H_{\rm el})'} \,Q\, |\phi_{\rm el}\rangle 
+ \langle\phi_{\rm el}| \,Q\, \frac{1}{({\cal E}_{\rm el}-H_{\rm el})'}
\stackrel{\leftrightarrow}{H_{\rm n}}|\phi_{\rm el}\rangle,
\label{30}
\end{align}
where for arbitrary $\psi_{\rm el}$ and $\psi'_{\rm el}$
\begin{equation}
\langle\psi'_{\rm el}|\!\!\stackrel{\leftrightarrow}{H_{\rm n}}\!\!|
\psi_{\rm el}\rangle
 = \langle\vec\nabla_R\,\psi'_{\rm el}|\vec\nabla_R\,\psi_{\rm  el}\rangle/(2\,m_{\rm n})
+\langle\psi'_{\rm el}|H''_{\rm n}|\psi_{\rm el}\rangle, \label{31}
\end{equation}
and where $m_{\rm n}$ is the reduced nuclear mass.

\section{Electric quadrupole moment}
The electric quadrupole moment operator is
\begin{align}
D^{ij} =&\ \sum_\alpha e_\alpha\,(r_{\alpha M}^i\,r_{\alpha M}^j- r_{\alpha M}^2\,\delta^{ij}/3)\,, \label{33}
\end{align}
where the summation index $\alpha$ refers to both electrons and nuclei, $e_{\alpha}$ refers to the electric charge of the $\alpha$-th particle,
and $\vec r_{\alpha M}$ are particle positions with respect to the center of mass, $\vec r_{M}$, 
\begin{align}
\vec r_{M} =&\ \sum_\alpha \frac{m_\alpha\,\vec r_\alpha}{M}\,. \label{34}
\end{align}
For the hydrogen molecule, $\sum_\alpha e_\alpha = 0$, and 
$M_A=M_B = m_N$, so the total mass $M$ is
\begin{align}
M =&\ 2\,m_N + 2\,m\,. \label{35}
\end{align}
Let us rewrite the quadrupole moment operator 
in terms of position vectors $\vec x_a = \vec r_{aG}$ with respect to the geometric center $\vec r_G$,
which coincides with the nuclear mass center.
\begin{align}
\vec r_G =&\ (\vec r_A + \vec r_B)/2. \label{36}
\end{align}
Namely
\begin{align}
\vec x_G =&\ \vec r_G -\vec r_{M}
=
 \sum_\alpha \frac{m_\alpha}{M} \,\vec r_G - \sum_\alpha \frac{m_\alpha\,\vec r_\alpha}{M}
=  
 - \sum_\alpha \frac{m_\alpha\,\vec r_{\alpha G}}{M}. \label{37}
\end{align}
Because $\vec r_G$ is the geometric center, the sum over nuclei cancels out, and
\begin{align}
\vec x_G =&\  - \frac{m}{M}\,\vec r_\mathrm{el}\,, \label{38}
\end{align}
where
\begin{align}
\vec r_\mathrm{el}\ = \sum_a \vec r_{aG}\,. \label{39}
\end{align}
The quadrupole moment operator in new variables becomes
\begin{align}
D^{ij} =&\ 
 \sum_\alpha e_a\bigg[(r_{\alpha G}^i + x_G^i)\,(r_{\alpha G}^j + x_G^j)- (\vec r_{\alpha G}+\vec x_G)^2\,\frac{\delta^{ij}}{3}\bigg]
\nonumber \\ =&\ 
D_G^{ij} + D_G^i\,x_G^j + x_G^i\,D_G^j - 2\,\vec D_G\,\vec x_G\,\frac{\delta^{ij}}{3} + O\bigg(\frac{m}{m_N}\bigg)^2 \label{40}
\end{align}
where, assuming $e$ is the electron charge, 
\begin{align}
D_G^i =&\ \sum_\alpha e_\alpha\,r_{\alpha G}^i = e\,r_\mathrm{el}^i \label{41}
\end{align}
and
\begin{align}
D_G^{ij} =&\  \sum_\alpha e_\alpha\,\bigg(r_{\alpha G}^i\,r_{\alpha G}^j- \vec r_{\alpha G}^{\,2}\,\frac{\delta^{ij}}{3}\bigg) 
= 
e\,r_\mathrm{el}^{ij} -  \frac{e}{2}\,\bigg(R^i\,R^j- \vec R^{\,2}\,\frac{\delta^{ij}}{3}\bigg) \label{42}
\end{align}
with
\begin{align}
r_\mathrm{el}^{ij} \equiv&\ \sum_a \bigg(r_{aG}^i\,r_{aG}^j- \vec r_{aG}^{\,2}\,\frac{\delta^{ij}}{3}\bigg). \label{43}
\end{align}
The quadrupole moment operator is therefore
\begin{align}
D^{ij} =&\ 
e\,r_\mathrm{el}^{ij} -  \frac{e}{2}\,\bigg(R^i\,R^j- \vec R^{\,2}\,\frac{\delta^{ij}}{3}\bigg)
- e\,\frac{m}{m_N}\,\bigg(r_\mathrm{el}^i\,r_\mathrm{el}^j -\vec r_\mathrm{el}^{\,2}\,\frac{\delta^{ij}}{3}\bigg) 
+ O\bigg(\frac{m}{m_N}\bigg)^2\,. \label{44}
\end{align}
Its matrix element on the $\phi_\mathrm{el}$ state in the BO approximation is
\begin{align}
D^{(0)ij}_\mathrm{el}=&\   e\,\bigg(n^i\,n^j- \frac{\delta^{ij}}{3}\bigg)\,D^{(0)}(R) \label{45}
\end{align}
where $n^i \equiv R^i/R$ and
\begin{align}
D^{(0)}(R) =&\ \frac{3}{2}\bigg[\big\langle r_\mathrm{el}^{ij}\,n^i\,n^j\big\rangle -\frac{R^2}{3}\bigg]\,, \label{46}
\end{align}
while the leading nonadiabatic correction is
\begin{align}
D^{(1)ij}_\mathrm{el}=&\ 
- e\,\frac{m}{m_N}\,\bigg\langle r_\mathrm{el}^i\,r_\mathrm{el}^j -\vec r_\mathrm{el}^{\,2}\,\frac{\delta^{ij}}{3} \bigg\rangle
+2\,e\,\bigg\langle r_\mathrm{el}^{ij}\,\frac{1}{({\cal E}_{\rm el}-H_{\rm el})'}\,
\stackrel{\leftrightarrow}{H_{\rm n}}\! \bigg\rangle. \label{47}
\end{align}
Let us denote
\begin{align}
|\phi_\mathrm{el}^{ij}\rangle = &\ 
\frac{1}{({\cal E}_{\rm el}-H_{\rm el})'}\,r_\mathrm{el}^{ij}|\phi_\mathrm{el}\rangle\,, \label{48}\\
 |\phi_\Pi^{k}\rangle = &\ 
  \frac{1}{({\cal E}_{\rm el}-H_{\rm el})}\,\sum_a (\vec n\cdot\vec r_a)\,r_{a\perp}^k|\phi_\mathrm{el}\rangle\,, \label{49} \\
|\psiS\rangle = &\ n^i\,n^j\, |\phi_\mathrm{el}^{ij}\rangle\,. \label{50}
\end{align}
Then, with reduced mass $m_\mathrm{n} = m_N/2$ the adiabatic correction to the quadrupole moment becomes
\begin{align}
D^{(1)ij}_\mathrm{el}=&\ 
 e\,\frac{m}{m_\mathrm{n}}\,\big(n^i\,n^j- \delta^{ij}/3\big)\,D^{(1)}(R)\,,  \label{51}
\end{align}
where 
\begin{align}
D^{(1)}(R) =&\ 
\frac{3}{2}\,\biggl[
-\frac{1}{2}\,\bigg\langle (\vec r_\mathrm{el}\cdot\vec n)^2 -\frac{\vec r_\mathrm{el}^{\,2}}{3}\bigg\rangle
+\frac{1}{m}\,\langle n^i\,n^j\,\nabla^k_R\phi_\mathrm{el}^{ij}|\nabla^k_R \phi_\mathrm{el}\rangle
+ \frac{1}{4\,m}\,\langle \psiS|  \vec p_\mathrm{el}^{\,2} |\phi_\mathrm{el}\rangle \biggr],  \label{52}
\end{align}
and where $\vec p_\mathrm{el}= \sum_a\vec q_a$.
Since
\begin{align}
n^i\,n^j\,\nabla^k =&\ 
-\frac{n^i}{R}\, \bigl(\delta^{kj}-n^k\,n^j)  - \frac{n^j}{R}\, \bigl(\delta^{ki}-n^k\,n^i) 
+ \nabla^k\,n^i\,n^j\,.  \label{53}
\end{align}
Then
\begin{align}
\langle n^i\,n^j\,\nabla^k_R\phi_\mathrm{el}^{ij}|\nabla^k_R \phi_\mathrm{el}\rangle =&\ 
\frac{1}{m}\langle \nabla^k_R\psiS|\nabla^k_R \phi_\mathrm{el}\rangle 
-\frac{2}{m\,R}\,\langle \phi_\Pi^{k} | \nabla^k_R \phi_\mathrm{el}\rangle\,. \label{54}
\end{align}
The last term in the above can be transformed using
\begin{equation}
\vec\nabla_R = \vec n\,(\vec n\cdot\vec\nabla_R) 
-\vec n\times(\vec n\times\vec\nabla_R),  \label{55}
\end{equation}
to the form
\begin{align}
\langle \vec \phi_\Pi | \vec \nabla_R \phi_\mathrm{el}\rangle =&\
-\langle \vec \phi_\Pi | \vec n\times(\vec n\times\vec\nabla_R) |\phi_\mathrm{el}\rangle
=
-\frac{i}{R}\,\langle \vec \phi_\Pi | \vec n\times\vec J_\mathrm{n} |\phi_\mathrm{el}\rangle
=
\frac{i}{R^2}\,\langle \vec \phi_\Pi | \vec R\times\vec J_\mathrm{el} |\phi_\mathrm{el}\rangle\,.  \label{56}
\end{align}
Thus, we have obtained the following formulas for the quadrupole moment in H$_2$ (in atomic units).
\begin{align}
D(R) =&\ D^{(0)}(R) +\frac{m}{m_\mathrm{n}}\,D^{(1)}(R) \label{57}\\
D^{(0)}(R) =&\ \frac{3}{2}\,Q_0(R)  \label{58}\\
D^{(1)}(R) =&\  \frac{3}{2}\,\big[Q_1(R)+Q_2(R)+Q_3(R)+Q_4(R)\big],  \label{59}
\end{align}
where
\begin{align}
Q_0(R)=&\ \sum_a \bigg\langle\phi_\mathrm{el}\bigg| (\vec n\cdot \vec r_a)^2 - \frac{\vec r_a^{\,2}}{3}\bigg|\phi_\mathrm{el} \bigg\rangle -\frac{R^2}{3},
 \label{60}\\
Q_1(R)=&\ -\frac{1}{2}\,\bigg\langle \phi_\mathrm{el}\bigg| (\vec r_\mathrm{el}\cdot\vec n)^2 -\frac{\vec r_\mathrm{el}^{\,2}}{3} \bigg|\phi_\mathrm{el}\bigg\rangle, 
\label{61}\\
Q_2(R)=&\ \langle \nabla^k_R\phi_{\Sigma^+} |\nabla^k_R \phi_\mathrm{el}\rangle\,, 
\label{62}\\
Q_3(R)=&\ -\frac{2\,i}{R^3}\langle \vec \phi_\Pi |\vec R\times\vec J_\mathrm{el}|\phi_\mathrm{el}\rangle, 
\label{63}\\
Q_4(R)=&\  \frac{1}{4}\,\langle \psiS|  \Big(\sum_a \vec p_a \Big)^2 |\phi_\mathrm{el}\rangle. \label{64}
\end{align}
This is the complete formula for the leading nonadiabatic correction to the quadrupole moment in a homonuclear diatomic molecule. The second term involves differentiation over the nuclear variable $\vec R$, and the following section describes a convenient way to calculate it with high precision.

\section{$R$-derivatives}
Following Ref. \cite{adjc}, let us now explain how one can calculate $R$-derivatives in the exponential basis.
Let $\psi_k$ be the $k$-th element of the basis set employed to expand
the ground-state electronic wave function 
\begin{equation}\label{65}
\phi_\mathrm{el} = \sum_k v_k\,\psi_k\,,
\end{equation} 
and let $\vec{v}$ be a vector consisting of real coefficients of this expansion.
The adiabatic correction can be written as
\begin{align} \label{66}
\bk{\nr\phi_\mathrm{el}}{\nr\phi_\mathrm{el}} =&\ \sum_{k,l} 
v_k\,v_l\, \bk{\nr\psi_k}{\nr\psi_l} 
+ \partial_R v_k\,\partial_R v_l\,\bk{\psi_k}{\psi_l} 
+ 2\,\partial_R v_k\,v_l\, \bk{\psi_k}{\partial_R\psi_l}\,,
\end{align}
where we assume that nonlinear parameters do not depend on $R$.
Next, let us define the following matrices
\begin{align}
{\cal H}_{kl} =&\ \bk{\psi_k}{\Hel\psi_l}\,,  \label{67}\\
{\cal N}_{kl} =&\ \bk{\psi_k}{\psi_l}\,, \label{68}\\
{\cal A}_{kl} =&\ \bk{\psi_k}{\partial_R\psi_l}\,, \label{69}\\
{\cal B}_{kl} =&\ \bk{\nr\psi_k}{\nr\psi_l}\,. \label{70}
\end{align}
With this notation, the electronic Schr{\"o}dinger equation can be written in the matrix form as
\begin{equation}
({\cal H} - \Eel\,{\cal N})\,\vec{v} = 0\,. \label{71}
\end{equation}
Let us further consider the first-order $R$-derivative of $\phiel$
\begin{equation}
\partial_R\phiel = \sum_k\left( \psi_k\,\partial_R v_k + v_k\,\partial_R\psi_k \right)\,.  \label{72}
\end{equation}
The term $\partial_R\psi_k$ is assumed to be known, as it is the derivative of 
a basis function at constant values of nonlinear parameters. 
The derivative $\partial_R v_k$ can be obtained by taking the derivative of Eq. (\ref{71}),
namely
\begin{equation} 
({\cal H} - \Eel\,{\cal N})\,\partial_R \vec{v} + 
(\partial_R{\cal H} - \partial_R\Eel\,{\cal N} - \Eel\,\partial_R{\cal N})\,\vec{v} = 0,  \label{73}
\end{equation}
so that
\begin{equation}
\partial_R \vec{v} = \frac{1}{(\Eel\,{\cal N} - {\cal H})'}\,
(\partial_R{\cal H} - \Eel\,\partial_R{\cal N})\,\vec{v}
-\frac{1}{2}\,\vec{v}\,\left(\vec{v}^T\;\partial_R {\cal N}\,\vec{v}\right)\,, \label{74}
\end{equation}
where the last term was obtained by differentiation of the normalization condition
\begin{equation}
\vec{v}^T\,{\cal N}\,\vec{v} = 1  \label{75}
\end{equation}
leading to
\begin{equation}
2\,(\partial_R\vec{v})^T\, {\cal N}\,\vec{v} + \vec{v}^T\,\partial_R {\cal N}\,\vec{v} = 0\,. \label{76}
\end{equation}
Now, the adiabatic correction of Eq.~(\ref{66}) is transformed to the form
\begin{equation} \label{77}
\langle \nabla^k_R\phi_\mathrm{el}|\nabla^k_R \phi_\mathrm{el}\rangle = 
\vec{v}^T\,{\cal B}\,\vec{v}
+ (\partial_R \vec{v})^T\,{\cal N}\,\partial_R \vec{v}
+ 2\,(\partial_R\vec{v})^T\,{\cal A}\,\vec{v}\,. 
\end{equation}

To calculate $Q_2$, which contains $R$-derivatives, we define
\begin{align}
\delta H =&\ \xi\,n^i\,n^j\,r^{ij}_\mathrm{el} \label{78}
\end{align}
and add it to the nonrelativistic Hamiltonian $H_\mathrm{el}$. So that $Q_2(R)$ can be obtained by 
a differentiation in $\xi$,
\begin{align}
Q_2 =&\ \frac{1}{2}\,\frac{d}{d\,\xi} \biggr|_{\xi=0}\,
\langle \nabla^k_R\phi_\mathrm{el}|\nabla^k_R \phi_\mathrm{el}\rangle_\xi\,,  \label{79}
\end{align}
where the subscript $\xi$ in $\langle\ldots\rangle_{\xi}$ stands for the $\xi$-dependence of $\phiel$ due to 
$\xi$-dependence of the Hamiltonian in Eq.~\eqref{78}. The methods for calculation of $Q_2$ are described in the next Section and in Appendix A.

\section{Numerical calculations}

\begin{table*}[t]
  \tbl{Numerical results for $D^{(0)}(R)=\tfrac{3}{2}Q_0(R)$ and $D^{(1)}(R)=\tfrac{3}{2}\sum_{i=1\ldots 4} Q_i(R)$. For $R<10$ au we used the single-sector JC basis ($y=x=0, \alpha=u=w$, $\Omega=15,16$), and for $R\ge10$ au a single-sector HL~basis ($\alpha=-y=x=u=w=1/2$, $\Omega=13,14$).}{

  \centering
  \begin{tabular*}{\textwidth}{@{\extracolsep{\fill}} 
    w{2.2}w{3.12}w{3.12}w{2.1}w{2.20}w{3.18}}
    \toprule
    \multicolumn{1}{c}{$R$} & \multicolumn{1}{c}{$D^{(0)}(R)$} & \multicolumn{1}{c}{$D^{(1)}(R)$} &
    \multicolumn{1}{c}{$R$} & \multicolumn{1}{c}{$D^{(0)}(R)$} & \multicolumn{1}{c}{$D^{(1)}(R)$} \\
    \midrule
0.1 & -0.002\,951\,435(4)	&-0.006\,320(8) & 4.6 & -0.424\,390\,384\,5(3)	&-1.467\,309\,361\,8(6) \\
0.2 & -0.011\,709\,44(1)	&-0.021\,260(9) & 4.8 & -0.350\,786\,848\,4(3)	&-1.188\,240\,220\,9(5) \\
0.3 & -0.026\,051\,81(1)	&-0.042\,022(2) & 5 & -0.286\,972\,211\,6(3)	&-0.953\,297\,753\,5(4) \\
0.35 & -0.035\,231\,298(6)	&-0.054\,283\,8(9) & 5.2 & -0.232\,780\,011\,8(3)	&-0.759\,883\,733\,8(4) \\
0.4 & -0.045\,699\,563(4)	&-0.067\,746\,1(4) & 5.4 & -0.187\,509\,569\,7(3)	&-0.602\,961\,001\,4(4) \\
0.45 & -0.057\,417\,112(3)	&-0.082\,401\,4(2) & 5.6 & -0.150\,184\,321\,3(3)	&-0.476\,861\,125\,0(4) \\
0.5 & -0.070\,343\,172(2)	&-0.098\,257\,1(1) & 5.8 & -0.119\,731\,937\,2(3)	&-0.376\,191\,601\,6(4) \\
0.55 & -0.084\,435\,808(2)	&-0.115\,330\,71(8) & 6 & -0.095\,097\,737\,3(2)	&-0.296\,203\,401\,3(4) \\
0.6 & -0.099\,651\,980(2)	&-0.133\,646\,81(6) & 6.5 & -0.052\,855\,949\,0(1)	&-0.162\,042\,133\,8(2) \\
0.65 & -0.115\,947\,571(2)	&-0.153\,234\,91(5) & 7 & -0.029\,133\,976\,98(9)	&-0.088\,399\,260\,3(1) \\
0.7 & -0.133\,277\,395(1)	&-0.174\,128\,40(4) & 7.5 & -0.016\,075\,011\,17(5)	&-0.048\,441\,307\,7(4) \\
0.75 & -0.151\,595\,187(1)	&-0.196\,363\,74(3) & 8 & -0.008\,963\,605\,49(3)	&-0.026\,898\,288\,7(7) \\
0.8 & -0.170\,853\,579\,1(9)	&-0.219\,979\,96(2) & 8.5 & -0.005\,101\,602\,00(2)	&-0.015\,283\,959(1) \\
0.85 & -0.191\,004\,076\,5(8)	&-0.245\,018\,41(1) & 9 & -0.002\,992\,484\,99(1)	&-0.008\,973\,903(1) \\
0.9 & -0.211\,997\,024\,8(7)	&-0.271\,522\,49(1) & 9.5 & -0.001\,823\,758\,72(1)	&-0.005\,487\,953\,2(1) \\
0.95 & -0.233\,781\,573\,5(6)	&-0.299\,537\,567(8) & 10 & -0.001\,160\,487\,87(7)	&-0.003\,511\,018\,5(2) \\
1 & -0.256\,305\,639\,0(5)	&-0.329\,110\,882(6) & 10.5 & -0.000\,771\,617\,62(5)	&-0.002\,349\,925\,1(2) \\
1.05 & -0.279\,515\,865\,2(5)	&-0.360\,291\,479(5) & 11 & -0.000\,534\,517\,89(3)	&-0.001\,639\,140\,3(1) \\
1.1 & -0.303\,357\,584\,4(5)	&-0.393\,130\,134(4) & 11.5 & -0.000\,383\,687\,60(2)	&-0.001\,184\,382\,29(9) \\
1.15 & -0.327\,774\,779\,1(4)	&-0.427\,679\,284(4) & 12 & -0.000\,283\,632\,66(1)	&-0.000\,880\,703\,22(8) \\
1.2 & -0.352\,710\,044\,2(4)	&-0.463\,992\,910(3) & 12.5 & -0.000\,214\,676\,34(1)	&-0.000\,669\,983\,92(6) \\
1.25 & -0.378\,104\,552\,1(4)	&-0.502\,126\,397(3) & 13 & -0.000\,165\,571\,539(7)	&-0.000\,518\,962\,39(5) \\
1.3 & -0.403\,898\,020\,3(4)	&-0.542\,136\,330(3) & 13.5 & -0.000\,129\,649\,273(5)	&-0.000\,407\,848\,30(4) \\
1.32 & -0.414\,313\,331\,5(4)	&-0.558\,678\,568(3) & 14 & -0.000\,102\,794\,279(3)	&-0.000\,324\,366\,36(3) \\
1.34 & -0.424\,778\,590\,6(4)	&-0.575\,533\,947(3) & 14.5 & -0.000\,082\,364\,189(2)	&-0.000\,260\,585\,51(2) \\
1.36 & -0.435\,289\,738\,8(4)	&-0.592\,706\,189(3) & 15 & -0.000\,066\,598\,895(2)	&-0.000\,211\,187\,81(1) \\
1.38 & -0.445\,842\,678\,0(4)	&-0.610\,199\,025(2) & 15.5 & -0.000\,054\,287\,879(1)	&-0.000\,172\,492\,27(1) \\
1.39 & -0.451\,133\,528\,1(4)	&-0.619\,066\,831(2) & 16 & -0.000\,044\,576\,061\,9(6)	&-0.000\,141\,883\,297(9) \\
1.4 & -0.456\,433\,271\,5(4)	&-0.628\,016\,185(2) & 16.5 & -0.000\,036\,846\,108\,9(4)	&-0.000\,117\,462\,538(8) \\
1.401\,1 & -0.457\,016\,763\,8(4)	&-0.629\,005\,612(2) & 17 & -0.000\,030\,644\,336\,2(2)	&-0.000\,097\,828\,413(8) \\
1.41 & -0.461\,741\,385\,1(4)	&-0.637\,047\,552(2) & 17.5 & -0.000\,025\,632\,422\,9(1)	&-0.000\,081\,931\,504(7) \\
1.42 & -0.467\,057\,343\,2(4)	&-0.646\,161\,396(2) & 18 & -0.000\,021\,554\,967\,77(8)	&-0.000\,068\,976\,789(6) \\
1.44 & -0.477\,710\,677\,7(4)	&-0.664\,638\,373(2) & 18.5 & -0.000\,018\,217\,128\,59(4)	&-0.000\,058\,355\,868(6) \\
1.46 & -0.488\,389\,020\,5(4)	&-0.683\,450\,812(2) & 19 & -0.000\,015\,468\,883\,22(2)	&-0.000\,049\,598\,999(6) \\
1.48 & -0.499\,088\,077\,7(4)	&-0.702\,602\,387(2) & 19.5 & -0.000\,013\,193\,758\,12(1)	&-0.000\,042\,340\,584(5) \\
1.5 & -0.509\,803\,516\,5(4)	&-0.722\,096\,737(2) & 20 & -0.000\,011\,300\,639\,904(7)	&-0.000\,036\,293\,994(5) \\
1.55 & -0.536\,634\,995\,1(4)	&-0.772\,355\,724(2) & 21 & -0.000\,008\,388\,192\,347(2)	&-0.000\,026\,977\,838(5) \\
1.6 & -0.563\,472\,053\,5(4)	&-0.824\,833\,971(2) & 22 & -0.000\,006\,316\,417\,375\,6(5)	&-0.000\,020\,338\,922(4) \\
1.65 & -0.590\,243\,824\,8(4)	&-0.879\,582\,963(2) & 23 & -0.000\,004\,818\,654\,241\,7(2)	&-0.000\,015\,532\,065(3) \\
1.7 & -0.616\,877\,988\,5(4)	&-0.936\,649\,816(2) & 24 & -0.000\,003\,719\,879\,686\,32(7)	&-0.000\,012\,001\,058(3) \\
1.8 & -0.669\,437\,307\,5(4)	&-1.057\,894\,508(2) & 25 & -0.000\,002\,902\,964\,190\,07(5)	&-0.000\,009\,372\,825(2) \\
1.9 & -0.720\,545\,075\,7(4)	&-1.188\,794\,210(2) & 26 & -0.000\,002\,288\,127\,320\,35(3)	&-0.000\,007\,392\,753(2) \\
2 & -0.769\,579\,093\,0(4)	&-1.329\,381\,366(2) & 27 & -0.000\,001\,820\,146\,001\,64(3)	&-0.000\,005\,884\,301(2) \\
2.1 & -0.815\,905\,362\,7(4)	&-1.479\,406\,343(2) & 28 & -0.000\,001\,460\,226\,595\,03(2)	&-0.000\,004\,723\,262(1) \\
2.2 & -0.858\,885\,741\,2(4)	&-1.638\,231\,196(2) & 29 & -0.000\,001\,180\,743\,679\,13(2)	&-0.000\,003\,821\,075(1) \\
2.3 & -0.897\,888\,100\,1(3)	&-1.804\,709\,949(2) & 30 & -0.000\,000\,961\,775\,941\,64(1)	&-0.000\,003\,113\,796(1) \\
2.4 & -0.932\,299\,240\,0(3)	&-1.977\,064\,939(2) & 31 & -0.000\,000\,788\,789\,220\,76(4)	&-0.000\,002\,554\,730(2) \\
2.5 & -0.961\,540\,620\,6(3)	&-2.152\,775\,178(2) & 32 & -0.000\,000\,651\,064\,123\,09(3)	&-0.000\,002\,109\,401(2) \\
2.6 & -0.985\,086\,710\,8(3)	&-2.328\,499\,368(2) & 33 & -0.000\,000\,540\,614\,443\,29(3)	&-0.000\,001\,752\,106(2) \\
2.7 & -1.002\,485\,402\,7(3)	&-2.500\,061\,558(2) & 34 & -0.000\,000\,451\,433\,779\,64(2)	&-0.000\,001\,463\,496(2) \\
2.8 & -1.013\,379\,503\,2(3)	&-2.662\,528\,726(2) & 35 & -0.000\,000\,378\,964\,564\,88(2)	&-0.000\,001\,228\,881(1) \\
2.9 & -1.017\,527\,860\,2(2)	&-2.810\,404\,164(2) & 36 & -0.000\,000\,319\,719\,719\,55(2)	&-0.000\,001\,037\,014(1) \\
3 & -1.014\,824\,274\,6(2)	&-2.937\,946\,270(2) & 37 & -0.000\,000\,271\,010\,266\,92(1)	&-0.000\,000\,879\,217(1) \\
3.1 & -1.005\,312\,098\,0(2)	&-3.039\,599\,290(2) & 38 & -0.000\,000\,230\,747\,325\,92(1)	&-0.000\,000\,748\,746(1) \\
3.2 & -0.989\,192\,424\,6(1)	&-3.110\,494\,365(2) & 39 & -0.000\,000\,197\,296\,853\,86(1)	&-0.000\,000\,640\,322(1) \\
3.3 & -0.966\,824\,133\,6(1)	&-3.146\,952\,952(2) & 40 & -0.000\,000\,169\,372\,165\,56(1)	&-0.000\,000\,549\,788(1) \\
3.4 & -0.938\,714\,739\,4(1)	&-3.146\,909\,544(1) & 41 & -0.000\,000\,145\,953\,755\,17(3)	&-0.000\,000\,473\,846(2) \\
3.5 & -0.905\,502\,003\,6(1)	&-3.110\,174\,370(1) & 42 & -0.000\,000\,126\,229\,022\,94(2)	&-0.000\,000\,409\,869(2) \\
3.6 & -0.867\,927\,383\,9(1)	&-3.038\,481\,851(1) & 43 & -0.000\,000\,109\,546\,633\,34(2)	&-0.000\,000\,355\,749(2) \\
3.7 & -0.826\,803\,440\,0(1)	&-2.935\,311\,834(1) & 44 & -0.000\,000\,095\,381\,712\,57(2)	&-0.000\,000\,309\,788(2) \\
3.8 & -0.782\,978\,067\,0(1)	&-2.805\,516\,542(1) & 45 & -0.000\,000\,083\,309\,135\,50(2)	&-0.000\,000\,270\,610(2) \\
3.9 & -0.737\,298\,724\,5(1)	&-2.654\,822\,813(1) & 46 & -0.000\,000\,072\,982\,892\,96(1)	&-0.000\,000\,237\,094(2) \\
4 & -0.690\,579\,643\,5(2)	&-2.489\,296\,550(1) & 47 & -0.000\,000\,064\,120\,059\,59(1)	&-0.000\,000\,208\,324(2) \\
4.2 & -0.596\,955\,200\,4(2)	&-2.136\,860\,879\,3(9) & 48 & -0.000\,000\,056\,488\,265\,29(1)	&-0.000\,000\,183\,546(1) \\
4.4 & -0.507\,085\,534\,9(3)	&-1.787\,644\,342\,5(7) & 50 & -0.000\,000\,044\,184\,093\,148(8)	&-0.000\,000\,143\,591(1)
    \\ \hline \hline
  \end{tabular*}}  \label{tab:D01}
\end{table*}

Following Ref. \cite{pachucki:10} we employ a variational approach to obtain $\phi_\mathrm{el}$ for the ground electronic $\Sigma^+$ state 
and utilize explicitly correlated exponential functions with polynomial dependence on interparticle distances of the form~\cite{Kolos_1966}
\begin{eqnarray}
\Phi_{\{n\}} &=&
	e^{-y\,\eta_1 -x\,\eta_2 -u\,\xi_1 -w\,\xi_2}
	r_{12}^{n_0} \, \eta_1^{n_1} \, \eta_2^{n_2} \, \xi_1^{n_3} \, \xi_2^{n_4},  \label{80}
	\label{kw}
\end{eqnarray}
where $\eta_i$ and $\xi_i$ are proportional to confocal elliptic coordinates and are given by $\eta_i=r_{iA}-r_{iB}$,
$\xi_i=r_{iA}+r_{iB}$, with $i$ enumerating electrons and real $y,x,u,w$ nonlinear parameters subject to variational minimization.
By $\{n\}$ we denote an ordered set of interparticle coordinate exponents, $(n_0,n_1,n_2,n_3,n_4)$, which are
conventionally restricted by a shell parameter $\Omega$,
\begin{equation}
	\sum_{j=0}^4 n_j \le \Omega\,. \label{81}
\end{equation}
If a symmetry restriction is imposed, the set of allowed $\{n\}$ is constrained even further for special values of nonlinear parameters. 
By construction, the trial functions depend on two-electron coordinates and account for the correlation via explicit dependence 
on the coordinate $r_{12}$.  The electronic  wavefunction is represented as
\begin{align}
	\Psi_{\Sigma^{+}} =&\ \sum_{\{n\}} v_{\{n\}} \hat{S}^{+}_{AB} \hat{S}^{+}_{12} \Phi_{\{n\}},  \label{82}\\
	\Psi_{\Pi} =&\ \sum_{\{n\}} v_{\{n\}} \hat{S}^{-}_{AB} \hat{S}^{+}_{12}\, r_{1\perp}^i\,\Phi_{\{n\}}\,, \label{83}
\end{align}
where $r_\perp^i = r^i-n^i\,\vec n\,\vec r$,
$\hat{S}^{\pm}_{AB} = 1 \pm P_{AB}$ and $P_{AB}$ permutes the nuclei $A$ and $B$, $\hat{S}^{\pm}_{12} = 1 \pm P_{12}$ and $P_{12}$ interchange the two electrons, and appropriate $\pm$ signs are chosen to fulfill the symmetry criteria for \textit{gerade}/\emph{ungerade} and \textit{singlet}/\textit{triplet} states. By solving the secular equation one obtains linear coefficients $v_{\{n\}}$. Such a form of wavefunction expansion is commonly referred to as the Ko{\l}os-Wolniewicz basis~\cite{Kolos_1966}.
In our calculations we use two specific cases of the KW basis, namely the symmetric James-Coolidge (JC) basis ($y=x=0,
\alpha=u=w$) for $R\leq10$ au and symmetric Heitler-London (HL) basis ($\alpha=-y=x=u=w$) for $R\geq10$ au.
Consequently, the resulting parametrization of our trial basis is extremely compact -- just a single nonlinear parameter $\alpha$ and the integer shell constraint $\Omega$.

Matrix elements with KW functions can be expressed as a linear combination of $f$-integrals with various sets of $\{n\}$, which are defined as
\begin{align}
f_{\{n\}}(R) =&\ R\,\int \frac{d^3 r_1}{4\,\pi}\,\int \frac{d^3 r_2}{4\,\pi}\,
\frac{e^{ -w_1\,r_{12} - u\,\xi_1 - w\,\xi_2
                    - y\,\eta_1 - x\,\eta_2}}{r_{1A}\,r_{1B}\,r_{2A}\,r_{2B}}                   
\,r_{12}^{n_0-1}  \eta_1 ^{n_1} \eta_2 ^{n_2} \xi_1
^{n_3} \xi_2 ^{n_4}. \label{84}
\end{align}
This is because all matrix elements can be expressed in terms of interparticle distances and their derivatives.
For example, $Q_0$ and $Q_1$ can be written as the following expectation values
\begin{align}
Q_0 =&\ -\frac{1}{6}\,\langle R^2 + r_{1A}^2 + r_{1B}^2 + r_{2A}^2 + r_{2B}^2\rangle
+ \frac{1}{4\,R^2}\,\langle (r_{1A}^2 - r_{1B}^2)^2 + (r_{2A}^2 - r_{2B}^2)^2\rangle\,,  \label{85}\\
Q_1 =&\ \frac{1}{8\,R^2}\,\langle(r_{1A}^2 - r_{1B}^2 + r_{2A}^2 - r_{2B}^2)^2\rangle
+  \frac{1}{6}\,\langle  r_{1A}^2 + r_{1B}^2 + r_{2A}^2 + r_{2B}^2 - R^2 - r_{12}^2\rangle.  \label{86}
\end{align}
The efficient recursive evaluation of $f$-integrals in arbitrary (extended) precision is described in
Refs.~\cite{h2solv,kw}. The numerical algebra part of the calculations is performed using parallel \texttt{HSL\_mp54} Cholesky solver~\cite{hsl,Hogg} 
adapted to quad-double arithmetic precision (64 decimal digits)~\cite{qd}.

The other operators $Q_2$, $Q_3$, and $Q_4$ are second-order matrix elements and thus require the construction of intermediate states,
\begin{align}
|\psiS\rangle = &\  
\frac{1}{({\cal E}_{\rm el}-H_{\rm el})'}\,r_\mathrm{el}^{ij} n^i\,n^j |\phi_\mathrm{el}\rangle\,,  \label{87} \\
|\phi_\Pi^{k}\rangle = &\ 
\frac{1}{({\cal E}_{\rm el}-H_{\rm el})'}\,\sum_a (\vec n\cdot\vec r_a)\,r_{a\perp}^k|\phi_\mathrm{el}\rangle\,,   \label{88}
\end{align}
which are obtained by solving a corresponding linear equation in the basis of $\Psi_{\Sigma^+}$ and $\Psi_\Pi$ functions, respectively. 
Those bases for intermediate states carry exactly the same set of nonlinear parameters as the external basis -- optimized for the energy of the ground electronic state.
This considerably reduces the number of integrals that have to be computed, and for the case of $|\psiS\rangle$ allows
us to reuse the basis already constructed for the external (ground) state, hence we have $\Omega_{\Sigma^+}=\Omega$.
Whereas for $|\phi_\Pi^{k}\rangle$ we choose $\Omega_\Pi=\Omega-2$, which curiously, due to different symmetry
restrictions for $\Sigma_g^+$ and $\Pi_g$, results in almost the same basis sizes, both in JC and HL cases.

Out of all $Q_i(R)$, the calculation of $Q_2$ in Eq.~\eqref{79} is the most difficult. The simplest way to evaluate it is to perform the differentiation in $\xi$ numerically,
\begin{align}
Q_2 \approx \frac{1}{2}\,
\frac{\langle \nabla^k_R\phi_\mathrm{el}|\nabla^k_R \phi_\mathrm{el}\rangle_\xi - \langle \nabla^k_R\phi_\mathrm{el}|\nabla^k_R \phi_\mathrm{el}\rangle_{-\xi}}{2\,\xi}
\end{align}
The matrix element $\langle \nabla^k_R\phi_\mathrm{el}|\nabla^k_R \phi_\mathrm{el}\rangle_\xi$ is constructed according to Eq.~\eqref{77}, 
which requires solving for a new eigenpair for each~$\xi$, due to the $\xi$-dependence of the Hamiltonian. 
Already with $\xi=10^{-5}$, this numerical method yields accurate results ($\delta\approx-2.5\cdot10^{-7}$ at $R=1.4011$ au) 
in comparison to the method based on analytical $\xi$-differentiation, described in Appendix A.
This confirms that the accuracy of this method is limited by the basis size, rather than the numerical differentiation at $\xi=0$.

\begin{figure}[!ht]
\includegraphics[width=\columnwidth]{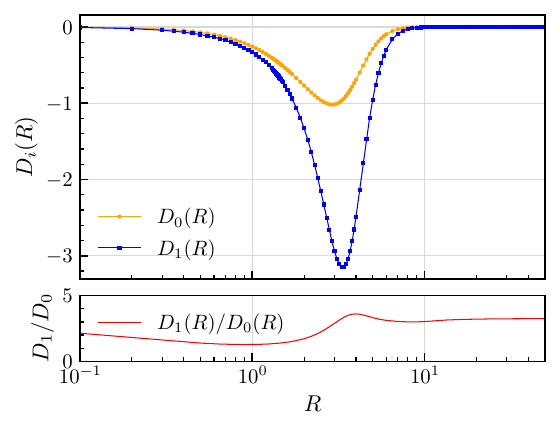}
\caption{Top: Born-Oppenheimer approximated quadrupole operator $D^{(0)}(R)$ (orange; circles) and its leading-order mass correction $D^{(1)}(R)$ (blue; squares). Bottom: Their ratio, supporting the claim that their leading-order long-range asymptotics is the same ($\sim\!R^{-6}$) up to a constant relative factor of about 3.25(5).}
\label{fig:D0D1}
\end{figure}

Numerical results for $D^{(0)}$ and $D^{(1)}$ are presented in Table \ref{tab:D01} and in Fig. \ref{fig:D0D1}.
These results have been obtained with a single sector JC basis ($\Omega=15,16$) for $R\leq 10$ and a HL basis 
($\Omega=13,14$) for $R\geq10$. Individual contributions $Q_i(R)$ to $D^{(0)}(R)$ and $D^{(1)}(R)$ are presented in 
Supplementary Material~\cite{supp}.
The relative accuracy of both $D^{(0)}(R)$ and $D^{(1)}(R)$ is typically no worse than $10^{-9}$ and becomes especially 
high ($\sim\! 10^{-17}$) for $D^{(0)}(R)$ at large $R$, proving that a simple single-sector HL basis is an excellent choice for this purpose.

As expected, the quadrupole moment and its finite nuclear mass correction vanish as $R\rightarrow 0$.
Even more important, and a stringent test of our calculations, is the long-range asymptotics. 
Although each of the individual parts, $Q_1$, $Q_2+Q_3$, and $Q_4$, has a leading $1/R^3$ term, those leading terms cancel out in the total sum.
Because the long-range asymptotics of $D^{(0)}(R)$ is $1/R^6$, see  Supplementary Material~\cite{supp}, one may expect a similar asymptotics for $D^{(1)}(R)$, and indeed this is the case; see the bottom panel of Fig. \ref{fig:D0D1}.

Comparing to the previous calculations, Wolniewicz~\cite{Wolniewicz98} defines $Q(R)$ by
\begin{align}
Q(R) =&\ 
\langle\phi_\mathrm{el}| \sum_a r_a^2\,\frac{3\,\cos^2(\theta_a)-1}{2} |\phi_\mathrm{el}\rangle\,.  \label{89}
\end{align}
Nevertheless, his numerical values correspond to the following definition
\begin{align}
Q(R) =&\ 
\langle\phi_\mathrm{el}| \sum_a r_a^2\,[1-3\,\cos^2(\theta_a)] |\phi_\mathrm{el}\rangle + R^2
=-2\,D^{(0)}(R)\,.  \label{90}
\end{align}
For example, his result for $R=1$ is $0.512\,611\,56$, which is nearly $-2$ times our result $D^{(0)}(1) = -0.256\,305\,639\,0(5)$. 
Komasa in Ref.~\cite{komasam1} defines $Q(R)$ with the opposite sign,
\begin{align}
Q(R) =&\ 
\langle\phi_\mathrm{el}| \sum_a r_a^2\,\frac{1-3\,\cos^2(\theta_a)}{2} |\phi_\mathrm{el}\rangle + \frac{R^2}{2} =-D^{(0)}(R)  \label{91}
\end{align}
and obtains $Q(1) = 0.256\,305\,641$.  We have found that aside from this prefactor, 
the numerical results of Wolniewicz \cite{Wolniewicz98} and Komasa \cite{komasam1} are in agreement with ours.

\section{Transition rates}
Following the notation from Ref. \cite{Wolniewicz98}, the spontaneous electric quadrupole transition probabilities 
from a (higher) initial state $v'J'$ to a (lower) final state $v''J''$, in atomic units, is given by (see Supplementary Material~\cite{supp})
\begin{align}
A_\mathrm{E2} =&\  (4\,\pi R_{\infty}c)\,\frac{\alpha^5}{15}\,(E_{v'J'}-E_{v''J''})^5 \, 
\langle v''J'' | D(R) | v'J' \rangle^2\,f(J',J'')\,,   \label{92}
\end{align}
where $R_{\infty}$ is the Rydberg constant, and rotational intensity factors are given by
\begin{align}
f(J',J'')= \left\{
\begin{array}{lr}
\dfrac{3(J'+1)(J'+2)}{2(2J'+1)(2J'+3)}, & J''=J'+2\ (\text{O})\\
\dfrac{J'(J'+1)}{(2J'-1)(2J'+3)},          & J''=J'\     (\text{Q})\\
\dfrac{3J'(J'-1)}{2(2J'-1)(2J'+1)},       & J''=J'-2\ (\text{S})
\end{array}\right.
\end{align}
and satisfy the identity
\begin{align}
f(J',J'+2) + f(J',J') + f(J',J'-2) = 1\,.
\end{align}
We note that Eq.\,(4) of Ref.~\cite{Wolniewicz98} contains a misprint in  the $J''=J'+2$ case, namely
the numerator should read $3(J'+1)(J'+2)$ instead of $(3J'+1)(J'+2)$.
Nevertheless, numerical transition rates in Ref.~\cite{Wolniewicz98} appear to be calculated with the correct rotational factors.

The radial matrix element in Eq. (\ref{92}) is calculated with $D(R)$  given by Eq. (\ref{57})
and with nuclear wavefunctions $\chi$  being a solution of the radial equation with the inclusion of the diagonal adiabatic correction.
We note that the $A_\mathrm{E2}$ rate is sensitive to this transition energy; therefore, 
we take values from the \textsc{H2Spectre} code~\cite{h2spectr}, which are accurate to about 1 MHz.
Our numerical results are in agreement with those of Wolniewicz~\cite{Wolniewicz98} and Komasa~\cite{komasam1}, 
but should be more accurate due to the inclusion of the leading nonadiabatic correction $D^{(1)}$.  
In fact, the E2 transition rates' uncertainty is now dominated by the unknown relativistic corrections, 
which we estimated by $\alpha^2$ times $A_\mathrm{E2}$.
We note, that magnetic dipole (M1) transitions occur at the same wavelengths as the electric quadrupole (E2) transitions.
Consequently, in the spontaneous emission the total transition probability is the sum of the separate M1 and E2 rates.
Therefore, whenever transition rates in the Q-branch ($J''=J'$) are of interest, one has to include also the M1 channel.
For the M1 transition rates in Table \ref{tab:rate}, we use Eq.\,(13) from Ref.~\cite{komasam1}, and our results are in a good agreement.

In Table~\ref{tab:rate}, apart from $A_\mathrm{M1}$,  we present the $A_{\mathrm{E2}}^Q$, $A_{\mathrm{E2}}^S$, and $A_{\mathrm{E2}}^O$ 
transition rates for the Q ($J''=J')$, S ($J''=J'-2$), and O ($J''=J'+2$) branches of the fundamental ($\nu=1\rightarrow 0$) band of H$_2$, respectively.

\begin{table}[!ht]
  \tbl{
    Transition rates for the Q ($J''=J')$, S ($J''=J'-2$), and O ($J''=J'+2$) branches of the fundamental ($\nu=1\rightarrow 0$) band of H$_2$ in units of $10^{-8}$~s$^{-1}$.
     $A_{\mathrm{M1}}$ is the magnetic dipole transition rate with the uncertainty due to unknown finite-nuclear-mass corrections estimated as $\sim 2\frac{m_e}{m_\mathrm{n}}A_{\mathrm{M1}}$.
     $\delta A_{\mathrm{E2}}^X$ is the difference in the E2 transition rate between those calculated with the mass-corrected quadrupole operator ($D^{(1)}$) with the adiabatically corrected nuclear wavefunction with $X=Q,S,O$, for Q-, S-, and O-branch, respectively; and the E2 rate in the BO approximation ($D^{(0)}$ operator and BO nuclear wavefunction). The E2 rates' uncertainty is now dominated by the unknown relativistic corrections, estimated by $\sim \alpha^2 A_\mathrm{E2} \approx 0.5 \cdot 10^{-4} A_\mathrm{E2}$. Physical constants are from Ref. \cite{codata2022}.}{
      \begin{tabular}{r w{2.7} w{2.5} w{1.2} w{2.6} w{1.2} w{2.8} w{1.2}}
\hline
      \mc{$J'$} & \mc{$A_{\mathrm{M1}}$} & \mc{$A_{\mathrm{E2}}^{Q}$} & \mc{$10^2\cdot\,\delta A_{\mathrm{E2}}^{Q}/A_{\mathrm{E2}}^{Q}$} & \mc{$A_{\mathrm{E2}}^{S}$} & \mc{$10^2\cdot\,\delta A_{\mathrm{E2}}^{S}/A_{\mathrm{E2}}^{S}$} & \mc{$A_{\mathrm{E2}}^{O}$} & \mc{$10^2\cdot\,\delta A_{\mathrm{E2}}^{O}/A_{\mathrm{E2}}^{O}$} \\
      \hline
0  & 0.0         & 0.0         & -    & 0.0         & -    & 85.601\,2   & 0.43 \\
1  & 0.071\,263  & 43.018\,58  & 0.44 & 0.0         & -    & 42.355\,57  & 0.43 \\
2  & 0.213\,856  & 30.394\,75  & 0.44 & 25.345\,57  & 0.45 & 29.048\,50  & 0.42 \\
3  & 0.427\,900  & 27.907\,32  & 0.44 & 34.821\,58  & 0.47 & 20.885\,10  & 0.42 \\
4  & 0.713\,535  & 26.592\,35  & 0.44 & 39.957\,98  & 0.48 & 15.022\,87  & 0.42 \\
5  & 1.070\,888  & 25.537\,08  & 0.45 & 42.228\,16  & 0.50 & 10.645\,77  & 0.41 \\
6  & 1.500\,002  & 24.520\,37  & 0.45 & 42.056\,38  & 0.52 & 7.379\,740  & 0.41 \\
7  & 2.000\,858  & 23.471\,48  & 0.45 & 39.717\,09  & 0.55 & 4.983\,036  & 0.41 \\
8  & 2.573\,176  & 22.367\,33  & 0.46 & 35.542\,90  & 0.59 & 3.266\,392  & 0.41 \\
9  & 3.216\,497  & 21.203\,88  & 0.46 & 29.974\,95  & 0.64 & 2.071\,560  & 0.41 \\
10 & 3.930\,060  & 19.985\,88  & 0.47 & 23.555\,94  & 0.71 & 1.266\,124  & 0.41 \\
11 & 4.712\,747  & 18.722\,64  & 0.47 & 16.897\,15  & 0.82 & 0.742\,0804 & 0.41 \\
12 & 5.563\,033  & 17.425\,88  & 0.48 & 10.634\,77  & 1.00 & 0.414\,3583 & 0.41 \\
13 & 6.478\,907  & 16.108\,52  & 0.49 & 5.384\,776  & 1.34 & 0.218\,4628 & 0.42 \\
14 & 7.457\,797  & 14.783\,94  & 0.50 & 1.702\,894  & 2.26 & 0.107\,4067 & 0.42 \\
15 & 8.496\,474  & 13.465\,42  & 0.51 & 0.053\,1505 & 11.79 & 0.048\,3604 & 0.42 \\
16 & 9.590\,935  & 12.165\,80  & 0.52 & 0.786\,6411 & -2.98 & 0.019\,4046 & 0.42 \\
17 & 10.736\,26    & 10.897\,15  & 0.53 & 4.130\,413  & -1.21 & 0.006\,6420 & 0.43 \\
18 & 11.926\,43    & 9.670\,58   & 0.55 & 10.185\,26  & -0.71 & 0.001\,7974 & 0.43 \\
19 & 13.154\,12    & 8.496\,08   & 0.56 & 18.930\,57  & -0.48 & 0.000\,3309 & 0.43 \\
20 & 14.410\,37    & 7.382\,42   & 0.58 & 30.234\,23  & -0.35 & 0.000\,0286 & 0.44 \\
21 & 15.684\,30    & 6.337\,12   & 0.60 & 43.865\,42  & -0.27 & 0.000\,000\,270 & 0.44 \\
22 & 16.962\,60    & 5.366\,34   & 0.63 & 59.508\,71  & -0.21 &             &      \\
23 & 18.228\,98    & 4.474\,97   & 0.66 & 76.777\,73  & -0.17 &             &      \\
24 & 19.463\,51    & 3.666\,56   & 0.69 & 95.227\,4   & -0.14 &             &      \\
25 & 20.641\,62    & 2.943\,40   & 0.73 & 114.363\,4  & -0.12 &             &      \\
26 & 21.732\,96    & 2.306\,50   & 0.78 & 133.648\,6  & -0.11 &             &      \\
27 & 22.699\,75    & 1.755\,61   & 0.84 & 152.503\,8  & -0.10 &             &      \\
28 & 23.494\,67    & 1.289\,29   & 0.92 & 170.303\,9  & -0.09 &             &      \\
29 & 24.057\,79    & 0.904\,82   & 1.01 & 186.363\,0  & -0.09 &             &      \\
30 & 24.311\,98    & 0.598\,28   & 1.14 & 199.905\,3  & -0.08 &             &      \\
\hline
    \end{tabular}}  \label{tab:rate}       
\end{table}

In the Q-branch, we note that except for $J'=1$, the magnetic dipole $A_\mathrm{M1}$~rate exceeds the nonadiabatic
corrections to $A_{\mathrm{E2}}^Q$, significantly grows with $J'$, and starting from $J'=18$ exceeds the total $A_{\mathrm{E2}}^Q$.
In the ($v = 1\rightarrow 0$) O-branch, the relative nonadiabatic correction is nearly constant and amounts to $\approx0.4\%$, the rapid drop in the magnitude of the overall E2 rate (as $J'$ is increased) in this branch can be attributed to the strong suppression by the $(E_{v'J'}-E_{v''J''})^5$ factor in Eq.~\eqref{92} as the energy difference goes to zero around $J'=21$.
Whereas for the S-branch, the relative nonadiabatic correction varies strongly, starting from $\approx0.45\%$ at $J'=2$, achieving a maximum of 12\% at $J'=15$ due to the smallness of the radial matrix element. This behaviour might be interesting to verify experimentally.  

\section{Summary}
We have derived formulas for nonadiabatic corrections to electric quadrupole transitions in H$_2$ using NAPT.
These corrections can be represented in terms of a single function $D^{(1)}(R)$, which is to be added to the BO function $D^{(0)}(R)$.
We have performed numerical calculations of $D^{(0)}$ and $D^{(1)}$ using James-Coolidge and Heitler-London basis functions. 
Results for $D^{(0)}$ are in agreement with previous ones~\cite{Wolniewicz98,komasam1}, but are much more accurate and in the much wider range $R\leq50$ a.u. 
Moreover, we have found  that the long-range asymptotics of  $D^{(1)}(R)$ is $\sim R^{-6}$, similarly to $D^{(0)}(R)$.

Using $D^{(0)}(R)$ and $D^{(1)}(R)$ curves, we have performed exemplary calculations of  
electric quadrupole transition rates in the $v=1\rightarrow0$ band for $J''=J',J'\pm 2$, and observed that nonadiabatic corrections to the Q- and O- branches are about 0.44\% - 1.14\%, 
which are a few to ten times larger than could be expected from a simple mass scaling by a factor of $m_e/m_\mathrm{n}$. 
In the S-branch at $J'=15$, where the rate is as small as $~5\cdot10^{-10}\,s^{-1}$, the nonadiabatic corrections raise the rate by as much as 12\%. 

With the improved quadrupole moment curve, one may recalculate all transition rates and lifetimes of H$_2$ (and D$_2$, T$_2$), which will be the subject of future work.
The ultimate way, however, is the direct nonadiabatic calculation currently being pursued by Komasa~\cite{directnonad}.
In addition, our results can be extended to heteronuclear molecules (such as HD) without major modifications. 
Moreover, we note that an analogous method can be used to derive formulas for nonadiabatic corrections for the electric polarizability and QED corrections to transition energies.

\section*{Acknowledgements}
We acknowledge interesting discussions with F. Merkt.

\section*{Disclosure statement}
No potential conflict of interest was reported by the authors.

\section*{Funding}
M.S. acknowledges support from the National Science Center (Poland) Grant No. 2019/34/E/ST4/00451.

\appendix
\section{Alternative Calculation of $Q_2$}

$Q_2(R)$ requires evaluation of $\langle \nabla^k_R\psiS|\nabla^k_R \phi_\mathrm{el}\rangle$.
Below we derive an explicit expression for Eq. \eqref{79}, starting from Eq.~\eqref{77},
\begin{align} \label{A1}
\langle \nabla^k_R\psiS|\nabla^k_R \phi_\mathrm{el}\rangle =&\
\frac{1}{2}\,\frac{d}{d\,\xi} \biggr|_{\xi=0}\,
[\vec{v}^T\,{\cal B}\,\vec{v}
+ (\partial_R \vec{v})^T\,{\cal N}\,\partial_R \vec{v}
+ 2\,(\partial_R\vec{v})^T\,{\cal A}\,\vec{v}]_\xi\,.
\end{align}
The matrices ${\cal B}, {\cal N}, {\cal A}$ do not depend on $\delta H$, so
\begin{align}
\langle \nabla^k_R\psiS|\nabla^k_R \phi_\mathrm{el}\rangle =&\ 
\vec{v}^T\,{\cal B}\,\delta\vec{v}
+ (\partial_R \vec{v})^T\,{\cal N}\,\partial_R \delta \vec{v}
+ (\partial_R \delta\vec{v})^T\,{\cal A}\,\vec{v} + (\partial_R\vec{v})^T\,{\cal A}\,\delta\vec{v}\,,
 \label{A2}
\end{align}
where
\begin{align}
\delta \vec v =&\ \frac{d \vec v}{d\xi}\biggr|_{\xi=0} \,.
 \label{A3}
\end{align}
$\delta \vec v $ can be  obtained by differentiation of the linear equation $({\cal H} -{\cal E}_\mathrm{el}\,{\cal N})\vec v = 0$ over $\xi$
\begin{align} \label{A4}
({\cal H} -{\cal E}_\mathrm{el}\,{\cal N})\delta \vec v + (\delta {\cal H} - \delta {\cal E}_\mathrm{el}\,{\cal N})\vec v = 0\,,
\end{align}
and by differentiation of the normalization condition $(\vec v^T{\cal N}\vec v)_\xi =1$ over $\xi$, so $\delta \vec v$ is orthogonal to $\vec v$,
namely $\vec v^T{\cal N}\delta \vec v=0$, and
\begin{align} \label{A5}
\delta \vec v =&\ \frac{1}{(\Eel\,{\cal N} - {\cal H})'}\,\delta {\cal H}\,\vec v\,,
\end{align}
where $\delta \mathcal{H}_{kl} \equiv \bk{\psi_k}{\delta H\big| \psi_l}$.
Next, we differentiate Eq. (\ref{73}) over $\xi$
\begin{align} \label{A6}
({\cal H} - \Eel\,{\cal N})\,\partial_R \delta \vec{v} + (\delta {\cal H} - \delta\Eel\,{\cal N})\,\partial_R \vec{v} 
+ \partial_R({\cal H} - \Eel\,{\cal N})\,\delta \vec{v} 
+ \partial_R\,(\delta {\cal H}  - \delta\Eel\,{\cal N})\,\vec{v} = 0\,,
\end{align}
to obtain
\begin{align} \label{A7}
\partial_R\delta \vec v =&\ 
\frac{1}{(\Eel\,{\cal N} - {\cal H})'}\,
\big[
\partial_R\,(\delta {\cal H}  - \delta\Eel\,{\cal N})\,\vec{v}
+ (\delta {\cal H} - \delta\Eel\,{\cal N})\,\partial_R \vec{v} 
+ \partial_R({\cal H} - \Eel\,{\cal N})\,\delta \vec{v} 
\big]\nonumber \\&
-\vec{v}\,\big[ (\partial_R\vec{v})^T\, {\cal N}\,\delta\vec{v} + \vec{v}^T\,\partial_R {\cal N}\,\delta\vec{v} \big]\,,
\end{align}
where the last term (parallel to $\vec v$) is found by differentiation of Eq. (\ref{76}) 
\begin{equation}
2\,(\partial_R\vec{v})^T\, {\cal N}\,\vec{v} + \vec{v}^T\,\partial_R {\cal N}\,\vec{v} = 0\,, \label{A8}
\end{equation}
over $\xi$
\begin{equation} \label{A9}
(\partial_R\delta\vec{v})^T\, {\cal N}\,\vec{v} 
+ (\partial_R\vec{v})^T\, {\cal N}\,\delta\vec{v} 
+ \vec{v}^T\,\partial_R {\cal N}\,\delta\vec{v} = 0\,.
\end{equation}
The derivative of ${\cal E}_\mathrm{el}$ over $R$ is obtained from
\begin{align} \label{A10}
0=&\ \partial_R [\vec v^T({\cal H} - \Eel\,{\cal N})\vec v] 
=
\vec v^T(\partial_R{\cal H} - \partial_R\Eel\,{\cal N} - \Eel\,\partial_R{\cal N})\vec v\,,
\end{align}
so
\begin{align}
\partial_R\Eel =&\ \vec v^T(\partial_R{\cal H}  - \Eel\,\partial_R{\cal N})\vec v \label{A11}
\end{align}
Similarly, the derivative of $\delta {\cal E}_\mathrm{el} = \vec v^T\delta{\cal H}\vec v$ over $R$ is obtained from
\begin{align} \label{A12}
\partial_R(\Eel + \delta \Eel) =&\ (\vec v^T + \delta\vec v^T)(\partial_R{\cal H}  + \partial_R\delta{\cal H} - (\Eel+\delta\Eel)\,\partial_R{\cal N})(\vec v +\delta\vec v) 
\end{align}
so
\begin{align} \label{A13}
\partial_R\delta\Eel =&\ \vec v^T(\partial_R\delta{\cal H}  - \delta\Eel\,\partial_R{\cal N})\vec v 
+ 2\,\vec v^T(\partial_R{\cal H} - \Eel\,\partial_R{\cal N})\delta\vec v
\end{align}
Summarizing, one calculates $Q_2$ using Eq. \eqref{A2}, with $\vec v$ obtained from Eq. \eqref{71}, $\partial_R \vec v$ from Eq. \eqref{74},
$\delta \vec v$ from Eq. (\ref{A5}), $\partial_R\delta \vec v$ from Eq. \eqref{A7}, $\partial_R\Eel$ from \eqref{A11}, and with $\partial_R\delta \Eel$ from \eqref{A13}.

\end{document}


\title{Supplementary material for:\\Nonadiabatic corrections to electric quadrupole transition rates in H$_2$}

\author{
\name{Krzysztof Pachucki\textsuperscript{a} and Michał Siłkowski\textsuperscript{b}}
\affil{\textsuperscript{a} Faculty of Physics, University of Warsaw, Pasteura 5, 02-093 Warsaw, Poland \\
\textsuperscript{b} Faculty of Chemistry, Adam Mickiewicz University in Pozna{\'n}, Uniwersytetu~Poznańskiego~8, 61-614 Pozna{\'n}, Poland}
}

\date{\today}

\maketitle

\section{Electric quadrupole transition rate}
We derive here the formula for the spontaneous electric quadrupole transition, starting from first principles.
The Fermi's golden rule for the transition rate is
\begin{align}
\frac{1}{\tau}=&\ \frac{2\,\pi}{\hbar^2}\,\int d\nu_f\,\big|\langle f| H_I |u\rangle\big|^2\,\delta(\omega_f-\omega_u),
\end{align}
where the integral is over all final states $f$, which have the same energy as the initial state $u$.

In the long-wavelength approximation, the interaction Hamiltonian is
\begin{align}
H_I = -\frac{1}{2}\,D^{ij}\,\nabla^jE^i(0)\,,
\end{align}
where $D^{ij}$ is the quadrupole moment operator, and $E^i_{,j}$ is the derivative of the electric field. 
The quantum electromagnetic field is
\begin{align}
\vec A(\vec r) =&\ \int\frac{d^3k}{\sqrt{(2\,\pi)^3}}\,\sqrt{\frac{\hbar}{2\,\epsilon_0\,\omega}}\,\vec\epsilon_{k,\lambda}\,
\big(a_{k,\lambda}\,e^{i\,\vec k\,\vec r} + a^+_{k,\lambda}\,e^{-i\,\vec k\,\vec r}\big)\,,\\
\vec E(\vec r)=&\ \int\frac{d^3k}{\sqrt{(2\,\pi)^3}}\,i\,\sqrt{\frac{\hbar\,\omega}{2\,\epsilon_0}}\,\vec\epsilon_{k,\lambda}\,
\big(a_{k,\lambda}\,e^{i\,\vec k\,\vec r} - a^+_{k,\lambda}\,e^{-i\,\vec k\,\vec r}\big)\,,
\end{align}
where
\begin{align}
[a_{k,\lambda}\,,\,a^+_{k',\lambda'}] =&\ \delta_{\lambda,\lambda'}\,\delta^3(\vec k-\vec k')\,.
\end{align}
Thus, the transition rate $A_\mathrm{E2} = 1/\tau$ with $\omega = k\,c$  is
\begin{align}
A_\mathrm{E2}=&\ 
\frac{2\,\pi}{\hbar^2}\,\int d^3k\,\sum_\lambda\,\Big\langle\vec k,\lambda; f\Big|\frac{1}{2}\,D^{ij}\,\nabla^jE^i(0)\Big| u; 0\Big\rangle^2\,\delta(\omega-\omega_0)
\nonumber \\ =&\
\frac{2\,\pi}{4\,\hbar^2}\,\frac{\hbar\,\omega_0}{2\,\epsilon_0}\,\int\frac{d^3k}{(2\,\pi)^3}\,\sum_\lambda\,
\Big|\langle f|D^{ij}|u\rangle\,\epsilon^i_{k,\lambda}\,k^j \Big|^2\,\delta(\omega-\omega_0)
\nonumber \\ =&\
\frac{2\,\pi}{4\,\hbar^2}\,\frac{\hbar\,\omega_0}{2\,\epsilon_0}\,\int\frac{d^3k}{(2\,\pi)^3}\,
\langle u| D^{mn}|f\rangle\,\langle f|D^{ij}|u\rangle\,
\sum_\lambda\,\epsilon^i_{k,\lambda}\,\epsilon^m_{k,\lambda}\,k^j\,k^n\,\delta(\omega-\omega_0)
\nonumber \\ =&\
\frac{\pi\,k_0}{4\,\hbar\,\epsilon_0}\,\int\frac{d^3k}{(2\,\pi)^3}\,
\langle u| D^{mn}|f\rangle\,\langle f|D^{ij}|u\rangle\,\biggl(\delta^{im}-\frac{k^i\,k^m}{k^2}\biggr)\,k^j\,k^n\,\delta(k-k_0)
\nonumber \\ =&\
\frac{k^5\,e^2}{8\,\pi\,\hbar\,\epsilon_0}\,
\langle u| d^{mn}|f\rangle\,\langle f|d^{ij}|u\rangle\,\int\frac{d\Omega}{4\,\pi}\,\bigl(\delta^{im}-n^i\,n^m\bigr)\,n^j\,n^n\,,
\end{align}
where $k=k_0$ from now on, $\vec n = \vec k/k$, and $D^{ij} = e\, d^{ij}$.
The angular integrations are
\begin{align}
\int\frac{d\Omega}{4\,\pi}\,n^j\,n^n =&\ \frac{\delta^{jn}}{3}\,,\\
\int\frac{d\Omega}{4\,\pi}\,n^i\,n^j\,n^m\,n^n =&\ \frac{1}{15}\,(\delta^{ij}\,\delta^{mn} + \delta^{im}\,\delta^{jn} + \delta^{in}\,\delta^{jm})\,.
\end{align}
Using $\alpha = e^2/(4\,\pi\,\epsilon_0\,\hbar\,c)$ the transition rate is
\begin{align}
A_\mathrm{E2}=&\ \alpha\,c\,k^5\,\langle u| d^{ij}|f\rangle\,\langle f|d^{ij}|u\rangle\,\frac{1}{10}
\end{align}
For a diatomic molecule the electric quadrupole operator can be written as
\begin{align}
d^{ij} = \bigg(n^i\,n^j-\frac{\delta^{ij}}{3}\bigg)\,D(R)\,
\end{align}
thus
\begin{align}
A_\mathrm{E2} =&\ 
\alpha\,c\,k^5\,\frac{1}{10}\,\langle \chi'' |D(R)| \chi'\rangle^2
\,\langle u| n^i\,n^j-\delta^{ij}/3|f\rangle\,\langle f| n^i\,n^j-\delta^{ij}/3|u\rangle
\end{align}
where $\chi', \chi''$ are initial and final radial functions, respectively.
\subsection*{Angular momentum algebra}
To calculate the rate for a specific angular momentum $J', J''$ of the initial and final state, respectively,
we make use of the angular momentum algebra,
\begin{align}
(n_1^i\,n_1^j -\delta^{ij}/3)\,(n_2^i\,n_2^j -\delta^{ij}/3) =&\  
\frac{2}{3}\,\sum_m C^*_{2m}(\theta_1,\phi_1)\,C_{2m}(\theta_2,\phi_2)
\end{align}
where
\begin{align}
C_{lm} =&\ \sqrt{\frac{4\,\pi}{2\,l+1}}\,Y_{lm}(\theta,\phi).
\end{align}
After averaging over initial $M'$ and summing over final $M''$ the rate is
\begin{align}
A_\mathrm{E2} =&\  \frac{1}{15}\,\alpha\,c\,k^5\,\langle v''J''|D(R)|v'J'\rangle^2\, X(J'')
\end{align}
where
\begin{align}
X(J'') =&\ \frac{1}{2\,J'+1}\sum_{M'}\,\sum_{M''}\,\sum_{q}\,|\langle J''M''|C_{2q}|J'M'\rangle|^2
\nonumber \\ =&\
(2\,J''+1)\,\threej{J'}{J''}{2}{0}{0}{0}^2
\nonumber \\ =&\
\left\{
\begin{array}{ll}
\frac{3\,(J'+1)\,(J'+2)}{2\,(2\,J'+1)\,(2\,J'+3)}   & \mathrm{for}\; J''=J'+2\,,\\[1ex]
\frac{J'\,(J'+1)}{(2\,J'-1)\,(2\,J'+3)} & \mathrm{for}\; J''=J'\,,\\[1ex]
\frac{3\,J'\,(J'-1)}{2\,(2\,J'-1)\,(2\,J'+1)} & \mathrm{for}\; J''=J'-2\,,
\end{array}
\right.
\end{align} 
and where the sum over all possible $J''$ is equal to 1,
\begin{align}
X(J'+2) + X(J') + X(J'-2) = 1\,.
\end{align}
Finally, the transition rate is
\begin{align}
A_\mathrm{E2} =&\ (4\,\pi R_\infty c)\, \frac{1}{15}\,\alpha^5\,\Delta E_\mathrm{au}^5
 \,X\,\langle v''J''|D(R)| v'J'\rangle^2_\mathrm{au}\,.
\end{align}

\newpage
\section{Detailed numerical results}

\begin{figure}[!b]
\centering

\subfloat[$Q_0(R)=\tfrac{2}{3}D_0(R)$ and $Q_i(R)$ — components of $D_1(R)$. The constant values of $Q_2(R)$ and $Q_3(R)$ cancel at $R\to0$ and $R\to\infty$.]{
  \includegraphics[width=0.46\linewidth]{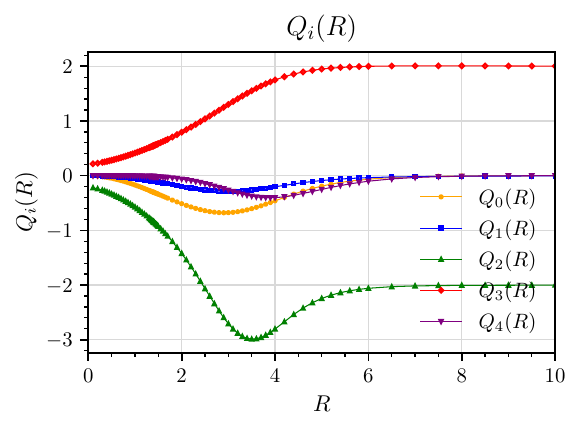}
  \label{fig:Q}
}\hfill
\subfloat[$Q_i(R)$ of $D_1(R)$ with $\sim R^{-3}$ tails: $R^3 Q_1(R)\!\to\!-4.33(2)$, $R^3(Q_2\!+\!Q_3)\!\to\!2.50(2)$, $R^3 Q_4(R)\!\to\!1.84(1)$; sum cancels (red).]{
  \includegraphics[width=0.46\linewidth]{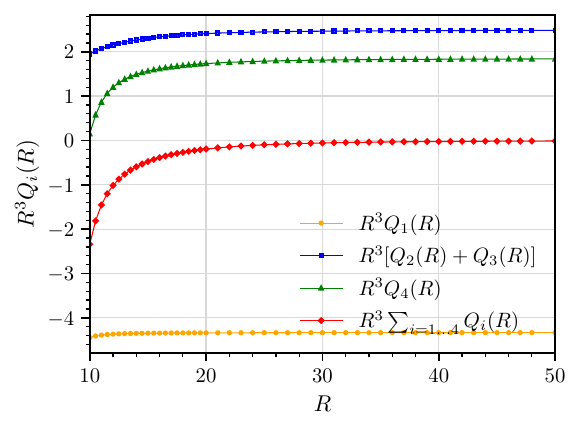}
  \label{fig:D1R3}
}

\medskip

\subfloat[$R^6 D_0(R)$]{
  \includegraphics[width=0.48\linewidth]{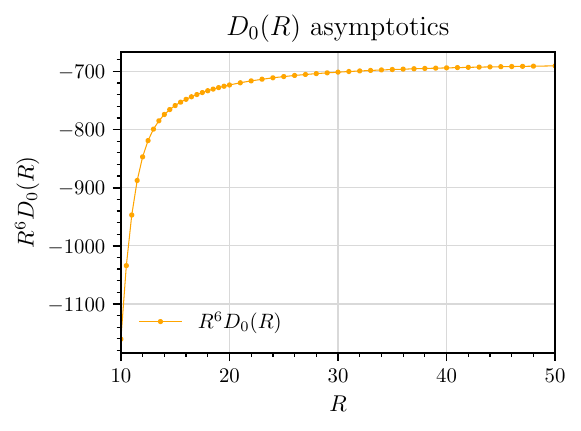}
  \label{fig:D0R6}
}\hfill
\subfloat[$R^6 D_1(R)$]{
  \includegraphics[width=0.48\linewidth]{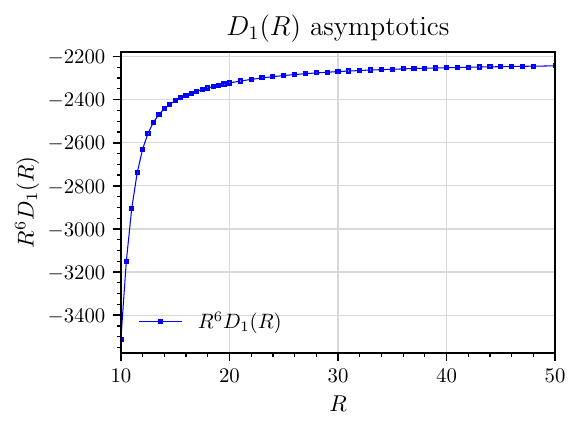}
  \label{fig:D1R6}
}

\caption{$R$-behavior of $D_0(R)$, $D_1(R)$ and their components $Q_i(R)$.}
\label{fig:2x2}
\end{figure}

\newpage

\begin{table}[t]
  \tbl{
  $Q_0(R)= \sum_a\langle (\vec r_a\cdot\vec n)^2 - r_a^2/3 \rangle - R^2/3$,
  $Q_1(R)= -\frac{1}{2}\langle\phi_\mathrm{el}| (\vec r_\mathrm{el}\cdot\vec n)^2 - \vec r_\mathrm{el}^{\,2}/3 |\phi_\mathrm{el}\rangle$,
  $Q_2(R)=\langle \nabla^k_R\psiS|\nabla^k_R \phi_\mathrm{el}\rangle$,
  $Q_3(R)= -2\,i/R^3\,\langle \vec\phi_\Pi | \vec R\times\vec J_\mathrm{el} |\phi_\mathrm{el}\rangle$,
  $Q_4(R)= \tfrac{1}{4}\langle \psiS| (\sum_a \vec p_a)^2 |\phi_\mathrm{el}\rangle$.
  Using single-sector JC ($y=x=0, \alpha=u=w$, $R\le10$~au, $\Omega=15,16$) and HL ($\alpha=-y=x=u=w=1/2$, $R>10$~au, $\Omega=13,14$) bases. $R_e=1.4011$~au. Nonlinear parameter $\alpha$ is dimensionless, all other quantities in atomic units.%
}{
  \centering
  \tiny
  \begin{tabular*}{\textwidth}{@{\extracolsep{\fill}} w{2.2}w{1.4}w{2.10}w{2.12}w{2.12}w{1.12}w{1.12}}
    \toprule
      \mc{$R$} & \mc{$\alpha$} & \mc{$Q^{(0)}(R)$} & \mc{$Q_1(R)$} & \mc{$Q_2(R)$} & \mc{$Q_3(R)$} & \mc{$Q_4(R)$} \\
    \midrule
    0.1	&1.7000	&-0.001967623(3)	&-0.0005376311(5)	&-0.220551(4)	&0.216839(4)	&0.0000360985(3)\\
0.2	&1.6790	&-0.007806294(10)	&-0.002155510(5)	&-0.240088(5)	&0.227949(4)	&0.00012155(4)\\
0.3	&1.5374	&-0.017367873(7)	&-0.004864231(4)	&-0.266631(2)	&0.2432698(10)	&0.00021047(2)\\
0.35	&1.4533	&-0.023487532(5)	&-0.006631002(3)	&-0.2819158(5)	&0.2521172(4)	&0.00024042(2)\\
0.4	&1.3693	&-0.030466375(3)	&-0.008674122(2)	&-0.2983707(2)	&0.2616287(2)	&0.000252051(7)\\
0.45	&1.3180	&-0.038278074(2)	&-0.0109942791(10)	&-0.3159175(2)	&0.27173878(10)	&0.000238711(6)\\
0.5	&1.2667	&-0.046895448(2)	&-0.0135917943(8)	&-0.33450472(6)	&0.28239819(6)	&0.000193605(4)\\
0.55	&1.2425	&-0.056290539(2)	&-0.0164665738(7)	&-0.35409949(4)	&0.29356917(4)	&0.000109751(4)\\
0.6	&1.2183	&-0.066434653(2)	&-0.0196180660(6)	&-0.37468238(3)	&0.30522263(4)	&-0.000020058(4)\\
0.65	&1.1983	&-0.077298381(2)	&-0.0230452214(6)	&-0.39624409(3)	&0.31733598(3)	&-0.000203278(3)\\
0.7	&1.1783	&-0.0888515970(10)	&-0.0267464531(5)	&-0.41878308(2)	&0.32989159(2)	&-0.000447650(3)\\
0.75	&1.1493	&-0.1010634581(8)	&-0.0307195984(4)	&-0.442303967(10)	&0.34287563(2)	&-0.000761224(3)\\
0.8	&1.1203	&-0.1139023861(7)	&-0.0349618804(4)	&-0.466816325(7)	&0.356277283(10)	&-0.001152387(2)\\
0.85	&1.0990	&-0.1273360510(6)	&-0.0394698694(3)	&-0.492333893(5)	&0.370088043(8)	&-0.001629889(2)\\
0.9	&1.0776	&-0.1413313498(5)	&-0.0442394434(3)	&-0.518873964(3)	&0.384301276(6)	&-0.002202863(2)\\
0.95	&1.0607	&-0.1558543824(5)	&-0.0492657486(3)	&-0.546456940(3)	&0.398911828(5)	&-0.002880850(2)\\
1.0	&1.0437	&-0.1708704260(4)	&-0.0545431590(2)	&-0.575106006(2)	&0.413915732(4)	&-0.003673822(2)\\
1.05	&1.0306	&-0.1863439101(4)	&-0.0600652357(2)	&-0.604846864(2)	&0.429309976(4)	&-0.0045921954(10)\\
1.1	&1.0174	&-0.2022383896(4)	&-0.0658246859(2)	&-0.6357075258(10)	&0.445092310(3)	&-0.0056468550(10)\\
1.15	&1.0070	&-0.2185165194(3)	&-0.0718133217(2)	&-0.6677181289(8)	&0.461261094(3)	&-0.0068491656(9)\\
1.2	&0.9965	&-0.2351400295(3)	&-0.0780220196(2)	&-0.7009107623(7)	&0.477815161(3)	&-0.0082109858(9)\\
1.25	&0.9876	&-0.2520697014(3)	&-0.0844406802(2)	&-0.7353192882(6)	&0.494753714(2)	&-0.0097446770(8)\\
1.3	&0.9794	&-0.2692653469(3)	&-0.0910581891(2)	&-0.7709791448(5)	&0.512076222(2)	&-0.0114631084(8)\\
1.32	&0.9763	&-0.2762088877(3)	&-0.0937581810(2)	&-0.7856016609(5)	&0.519112651(2)	&-0.0122051876(8)\\
1.34	&0.9734	&-0.2831857271(3)	&-0.0964872273(2)	&-0.8004326702(5)	&0.526210438(2)	&-0.0129798385(8)\\
1.36	&0.9705	&-0.2901931592(3)	&-0.0992444727(2)	&-0.8154746040(5)	&0.533369570(2)	&-0.0137879525(8)\\
1.38	&0.9677	&-0.2972284520(3)	&-0.1020290335(2)	&-0.8307299145(5)	&0.540590032(2)	&-0.0146304337(8)\\
1.39	&0.9663	&-0.3007556854(3)	&-0.1034312733(2)	&-0.8384383573(5)	&0.544223257(2)	&-0.0150648478(7)\\
1.4	&0.9650	&-0.3042888477(3)	&-0.1048399974(2)	&-0.8462010715(5)	&0.547871810(2)	&-0.0155081982(7)\\
\mc{$R_e$}	&0.9650	&-0.3046778425(3)	&-0.1049953479(2)	&-0.8470582967(5)	&0.548274087(2)	&-0.0155575173(7)\\
1.41	&0.9637	&-0.3078275901(3)	&-0.1062550874(2)	&-0.8540183680(5)	&0.551535689(2)	&-0.0159606015(7)\\
1.42	&0.9621	&-0.3113715621(3)	&-0.1076764227(2)	&-0.8618905581(5)	&0.555214892(2)	&-0.0164221748(7)\\
1.44	&0.9596	&-0.3184737851(3)	&-0.1105373382(2)	&-0.8778008665(5)	&0.562619260(2)	&-0.0173733036(7)\\
1.46	&0.9570	&-0.3255926803(3)	&-0.1134217434(2)	&-0.8939344933(5)	&0.570084899(2)	&-0.0183625367(7)\\
1.48	&0.9546	&-0.3327253851(3)	&-0.1163286074(2)	&-0.9102939343(5)	&0.577611787(2)	&-0.0193908373(7)\\
1.5	&0.9524	&-0.3398690110(3)	&-0.1192568694(2)	&-0.9268816792(5)	&0.585199904(2)	&-0.0204591794(7)\\
1.55	&0.9466	&-0.3577566634(3)	&-0.1266639052(2)	&-0.9693660835(5)	&0.604437872(2)	&-0.0233116993(7)\\
1.6	&0.9409	&-0.3756480357(3)	&-0.1341798887(2)	&-1.0133308866(5)	&0.6240577539(10)	&-0.0264362929(7)\\
1.65	&0.9360	&-0.3934958832(3)	&-0.1417857295(2)	&-1.0588127107(6)	&0.6440587139(10)	&-0.0298489155(7)\\
1.7	&0.9311	&-0.4112519924(3)	&-0.1494611292(2)	&-1.1058458972(6)	&0.6644395791(9)	&-0.0335657637(7)\\
1.8	&0.9218	&-0.4462915383(3)	&-0.1649334114(2)	&-1.2046862413(7)	&0.7063341658(8)	&-0.0419775182(7)\\
1.9	&0.9132	&-0.4803633838(3)	&-0.1804106351(2)	&-1.3100394357(8)	&0.7497222606(8)	&-0.0518016635(7)\\
2.0	&0.9051	&-0.5130527287(3)	&-0.1956888631(2)	&-1.4219756330(9)	&0.7945736713(8)	&-0.0631634192(7)\\
2.1	&0.8975	&-0.5439369085(3)	&-0.2105483388(2)	&-1.5403903886(10)	&0.8408440359(7)	&-0.0761762036(7)\\
2.2	&0.8900	&-0.5725904941(3)	&-0.2247566212(2)	&-1.664936137(2)	&0.8884712721(7)	&-0.0909326440(7)\\
2.3	&0.8829	&-0.5985920667(3)	&-0.2380729945(2)	&-1.794945313(2)	&0.9373717771(7)	&-0.1074934351(7)\\
2.4	&0.8759	&-0.6215328267(3)	&-0.2502542553(3)	&-1.929351346(2)	&0.9874365285(7)	&-0.1258742207(7)\\
2.5	&0.8690	&-0.6410270804(3)	&-0.2610618882(3)	&-2.066617833(2)	&1.0385273249(7)	&-0.1460310558(7)\\
2.6	&0.8623	&-0.6567244739(3)	&-0.2702705065(3)	&-2.204690419(2)	&1.0904735053(7)	&-0.1678454914(8)\\
2.7	&0.8556	&-0.6683236018(2)	&-0.2776772612(3)	&-2.340989142(2)	&1.1430695770(7)	&-0.1911108794(8)\\
2.8	&0.8489	&-0.6755863355(2)	&-0.2831117245(3)	&-2.472459657(2)	&1.1960742477(7)	&-0.2155220170(8)\\
2.9	&0.8422	&-0.6783519068(2)	&-0.2864455524(3)	&-2.5956980023(10)	&1.2492113733(6)	&-0.2406705948(8)\\
3.0	&0.8355	&-0.6765495164(2)	&-0.2876010713(3)	&-2.7071541490(9)	&1.3021732696(6)	&-0.2660488961(8)\\
3.1	&0.8286	&-0.6702080653(2)	&-0.2865578558(3)	&-2.8034046861(8)	&1.3546266779(6)	&-0.2910636624(8)\\
3.2	&0.8219	&-0.65946161642(9)	&-0.2833564101(3)	&-2.8814670114(7)	&1.4062214161(5)	&-0.3150609046(8)\\
3.3	&0.8147	&-0.64454942242(6)	&-0.2780982705(3)	&-2.9391109916(6)	&1.4566014158(5)	&-0.3373607886(8)\\
3.4	&0.8077	&-0.62580982628(4)	&-0.2709421893(3)	&-2.9751151531(5)	&1.5054174782(4)	&-0.3572998315(8)\\
3.5	&0.8004	&-0.603668002410(4)	&-0.2620965179(3)	&-2.9894178898(5)	&1.5523407705(4)	&-0.3742759427(7)\\
3.6	&0.7931	&-0.57861825589(3)	&-0.2518083808(3)	&-2.9831312404(5)	&1.5970758745(3)	&-0.3877908204(7)\\
3.7	&0.7852	&-0.55120229333(5)	&-0.2403506469(3)	&-2.9584118697(4)	&1.6393721811(2)	&-0.3974842207(7)\\
3.8	&0.7773	&-0.52198537799(8)	&-0.2280079590(3)	&-2.9182132369(4)	&1.6790325828(2)	&-0.4031557483(6)\\
3.9	&0.7692	&-0.49153248297(10)	&-0.2150631351(3)	&-2.8659656712(4)	&1.71591875660(10)	&-0.4047718258(6)\\
4.0	&0.7605	&-0.4603864290(2)	&-0.2017851030(3)	&-2.8052408030(4)	&1.74995275707(6)	&-0.4024578843(5)\\
4.2	&0.7432	&-0.3979701336(2)	&-0.1751804343(3)	&-2.6716274308(4)	&1.80943975219(2)	&-0.3872058067(4)\\
4.4	&0.7248	&-0.3380570232(2)	&-0.1497710691(3)	&-2.5393026299(3)	&1.85793560053(6)	&-0.3606247965(3)\\

    \\ \hline \hline
  \end{tabular*}}  \label{tab:Qlong}
\end{table}

\newpage

\begin{table}[t]
      \ContinuedFloat
  \tbl{(continued)}{
  \centering
  \tiny
  \begin{tabular*}{\textwidth}{w{1.1}w{0.2}w{2.18}w{2.15}w{2.12}w{2.14}w{2.12}}
    \toprule
      \mc{$R$} & \mc{$\alpha$} & \mc{$Q^{(0)}(R)$} & \mc{$Q_1(R)$} & \mc{$Q_2(R)$} & \mc{$Q_3(R)$} & \mc{$Q_4(R)$} \\
    \midrule
    4.6	&0.7057	&-0.2829269230(2)	&-0.1265823608(3)	&-2.4214706388(2)	&1.89648455854(8)	&-0.3266378000(2)\\
4.8	&0.6856	&-0.2338578990(2)	&-0.1061354647(3)	&-2.32349318854(8)	&1.92645889578(8)	&-0.28899038987(9)\\
5.0	&0.6653	&-0.1913148077(2)	&-0.0885653233(3)	&-2.245559261384(7)	&1.94932299928(8)	&-0.25073025029(2)\\
5.2	&0.6448	&-0.1551866745(2)	&-0.0737522640(3)	&-2.18527476597(7)	&1.96646989915(8)	&-0.21403202507(4)\\
5.4	&0.6243	&-0.1250063798(2)	&-0.0614342982(3)	&-2.13941814898(10)	&1.97913038768(7)	&-0.18025194143(7)\\
5.6	&0.6043	&-0.1001228809(2)	&-0.0512887177(3)	&-2.1048668952(2)	&1.98833796473(7)	&-0.15008976851(8)\\
5.8	&0.5857	&-0.0798212915(2)	&-0.0429844040(3)	&-2.0789662755(2)	&1.99492945218(6)	&-0.12377317371(8)\\
6.0	&0.5705	&-0.0633984915(2)	&-0.0362114176(2)	&-2.0596014782(2)	&1.99956458739(5)	&-0.10122062579(7)\\
6.5	&0.5631	&-0.03523729936(10)	&-0.0242732607(2)	&-2.03004328882(4)	&2.00564580210(3)	&-0.05935734181(2)\\
7.0	&0.5483	&-0.01942265132(6)	&-0.01705305479(8)	&-2.01586181279(5)	&2.007485677919(9)	&-0.03350365051(5)\\
7.5	&0.5350	&-0.01071667411(4)	&-0.01256361842(5)	&-2.0090507868(2)	&2.007540776661(2)	&-0.0182205766(2)\\
8.0	&0.5305	&-0.00597573699(2)	&-0.00965698897(3)	&-2.0057272937(4)	&2.006928682731(5)	&-0.0094765925(4)\\
8.5	&0.5300	&-0.00340106800(2)	&-0.00768496278(2)	&-2.0040303179(5)	&2.00613328030(2)	&-0.0046073056(5)\\
9.0	&0.5301	&-0.001994989991(7)	&-0.00628254793(2)	&-2.0030872750(5)	&2.00534978279(2)	&-0.0019625620(5)\\
9.5	&0.5302	&-0.001215839150(7)	&-0.00524184984(2)	&-2.00249969182(5)	&2.00464663129(4)	&-0.000563725073(7)\\
10.0	&0.5000	&-0.00077365858(5)	&-0.00444163471(8)	&-2.00208963536(3)	&2.00403922544(3)	&0.0001513657(2)\\
10.5	&0.5000	&-0.00051441175(4)	&-0.00380877235(6)	&-2.00177815561(3)	&2.00352255074(3)	&0.00049776051(9)\\
11.0	&0.5000	&-0.00035634526(3)	&-0.00329728647(4)	&-2.00152920409(2)	&2.00308510428(2)	&0.00064862608(7)\\
11.5	&0.5000	&-0.00025579173(2)	&-0.00287694618(3)	&-2.00132485423(2)	&2.002714510001(10)	&0.00069770221(6)\\
12.0	&0.5000	&-0.000189088440(10)	&-0.00252698135(3)	&-2.00115483236(2)	&2.002399545393(6)	&0.00069513284(5)\\
12.5	&0.5000	&-0.000143117560(7)	&-0.00223254784(2)	&-2.00101227838(2)	&2.002130682239(4)	&0.00066748804(4)\\
13.0	&0.5000	&-0.000110381026(5)	&-0.001982679342(10)	&-2.000892078027(9)	&2.001900055967(2)	&0.00062872648(4)\\
13.5	&0.5000	&-0.000086432849(4)	&-0.001769055669(7)	&-2.000790197595(6)	&2.001701248360(2)	&0.00058610604(3)\\
14.0	&0.5000	&-0.000068529520(3)	&-0.001585230565(5)	&-2.000703387411(4)	&2.0015290371430(6)	&0.00054333660(2)\\
14.5	&0.5000	&-0.000054909459(2)	&-0.001426126617(3)	&-2.000629015635(3)	&2.0013791675873(4)	&0.00050225099(2)\\
15.0	&0.5000	&-0.000044399263(2)	&-0.001287691560(2)	&-2.000564950406(2)	&2.0012481611432(2)	&0.000463688948(9)\\
15.5	&0.5000	&-0.0000361919193(7)	&-0.001166656207(2)	&-2.000509465075(3)	&2.00113316088668(8)	&0.000427965548(7)\\
16.0	&0.5000	&-0.0000297173746(5)	&-0.0010603589906(8)	&-2.000461159531(3)	&2.00103180867949(4)	&0.000395120977(6)\\
16.5	&0.5000	&-0.0000245640726(3)	&-0.0009666158482(5)	&-2.000418895326(3)	&2.00094214809297(2)	&0.000365054722(5)\\
17.0	&0.5000	&-0.0000204295575(2)	&-0.0008836219422(3)	&-2.000381743044(3)	&2.00086254772184(2)	&0.000337598322(5)\\
17.5	&0.5000	&-0.00001708828194(9)	&-0.0008098763214(2)	&-2.000348940188(3)	&2.000791640456261(7)	&0.000312555051(4)\\
18.0	&0.5000	&-0.00001436997851(6)	&-0.00074412347436(9)	&-2.000319857803(3)	&2.000728275201897(5)	&0.000289721550(4)\\
18.5	&0.5000	&-0.00001214475239(3)	&-0.00068530752270(6)	&-2.000293974157(3)	&2.000671478316788(4)	&0.000268899451(3)\\
19.0	&0.5000	&-0.00001031258881(2)	&-0.00063253600713(4)	&-2.000270854006(3)	&2.000620422655556(4)	&0.000249901359(3)\\
19.5	&0.5000	&-0.000008795838746(9)	&-0.00058505103452(3)	&-2.000250132247(3)	&2.000574402593992(3)	&0.000232553632(3)\\
20.0	&0.5000	&-0.000007533759936(5)	&-0.00054220612930(2)	&-2.000231500973(3)	&2.000532813777115(3)	&0.000216697329(3)\\
21.0	&0.5000	&-0.000005592128231(2)	&-0.00046829906954(2)	&-2.000199504583(3)	&2.000460922778439(3)	&0.000188895648(3)\\
22.0	&0.5000	&-0.0000042109449171(4)	&-0.000407243200031(9)	&-2.000173202770(2)	&2.000401383256083(3)	&0.000165503433(2)\\
23.0	&0.5000	&-0.0000032124361611(2)	&-0.000356360699981(7)	&-2.000151369465(2)	&2.000351656721260(2)	&0.000145718733(2)\\
24.0	&0.5000	&-0.00000247991979088(5)	&-0.000313616847844(7)	&-2.000133085500(2)	&2.000309804899576(2)	&0.000128896743(2)\\
25.0	&0.5000	&-0.00000193530946005(4)	&-0.000277446662120(6)	&-2.000117651754(2)	&2.000274330737534(2)	&0.000114519128(2)\\
26.0	&0.5000	&-0.00000152541821356(3)	&-0.000246633009547(5)	&-2.0001045295090(9)	&2.000244065696308(2)	&0.0001021683200(9)\\
27.0	&0.5000	&-0.00000121343066776(2)	&-0.000220219590512(5)	&-2.0000932987429(8)	&2.000218088735741(2)	&0.0000915067307(8)\\
28.0	&0.5000	&-0.00000097348439669(2)	&-0.000197447954389(4)	&-2.0000836284620(7)	&2.000195667321900(2)	&0.0000822602529(7)\\
29.0	&0.5000	&-0.00000078716245275(2)	&-0.000177711323972(4)	&-2.0000752552788(6)	&2.000176213954392(2)	&0.0000742052654(6)\\
30.0	&0.5000	&-0.00000064118396110(1)	&-0.000160520342875(4)	&-2.0000679677258(5)	&2.000159253772464(1)	&0.0000671584319(5)\\
31.0	&0.5000	&-0.00000052585948050(3)	&-0.000145477389339(9)	&-2.0000615946217(10)	&2.000144400165470(3)	&0.0000609686923(10)\\
32.0	&0.5000	&-0.00000043404274873(3)	&-0.000132257118553(9)	&-2.0000559963416(9)	&2.000131336231673(3)	&0.0000555109609(9)\\
33.0	&0.5000	&-0.00000036040962886(2)	&-0.000120591583835(8)	&-2.0000510581990(9)	&2.000119800555210(3)	&0.0000506811571(9)\\
34.0	&0.5000	&-0.00000030095585309(2)	&-0.000110258758884(7)	&-2.0000466853737(8)	&2.000109576202840(2)	&0.0000463922660(8)\\
35.0	&0.5000	&-0.00000025264304325(2)	&-0.000101073610573(7)	&-2.0000427989922(7)	&2.000100482143740(2)	&0.0000425712053(7)\\
36.0	&0.5000	&-0.00000021314647970(2)	&-0.000092881101811(6)	&-2.0000393330719(7)	&2.000092366508615(2)	&0.0000391563227(7)\\
37.0	&0.5000	&-0.000000180673511282(9)	&-0.000085550667307(6)	&-2.0000362321188(6)	&2.000085101256434(2)	&0.0000360953851(6)\\
38.0	&0.5000	&-0.000000153831550612(8)	&-0.000078971822339(5)	&-2.0000334492264(6)	&2.000078577926771(2)	&0.0000333439580(6)\\
39.0	&0.5000	&-0.000000131531235905(7)	&-0.000073050649606(5)	&-2.0000309445592(5)	&2.000072704235456(2)	&0.0000308640918(5)\\
40.0	&0.5000	&-0.000000112914777041(6)	&-0.000067706971432(5)	&-2.0000286841362(5)	&2.000067401329878(2)	&0.0000286232527(5)\\
41.0	&0.5000	&-0.00000009730250345(2)	&-0.00006287206048(2)	&-2.000026638849(2)	&2.000062601563609(4)	&0.000026593448(2)\\
42.0	&0.5000	&-0.00000008415268196(2)	&-0.00005848677629(2)	&-2.000024783661(2)	&2.000058246682442(4)	&0.000024750509(2)\\
43.0	&0.5000	&-0.00000007303108889(2)	&-0.00005450004051(2)	&-2.000023096962(1)	&2.000054286338229(3)	&0.0000230734976(10)\\
44.0	&0.5000	&-0.00000006358780838(2)	&-0.00005086758319(2)	&-2.000021560025(1)	&2.000050676865407(3)	&0.0000215442170(10)\\
45.0	&0.5000	&-0.00000005553942367(2)	&-0.00004755090699(2)	&-2.0000201565734(9)	&2.000047380269118(3)	&0.0000201468044(9)\\
46.0	&0.5000	&-0.00000004865526197(2)	&-0.000044516427673(10)	&-2.0000188724095(9)	&2.000044363384642(3)	&0.0000188673899(8)\\
47.0	&0.5000	&-0.000000042746706392(9)	&-0.000041734757848(9)	&-2.0000176951126(8)	&2.000041597176218(3)	&0.0000176938118(8)\\
48.0	&0.5000	&-0.000000037658843526(8)	&-0.000039180107555(9)	&-2.0000166137839(7)	&2.000039056149774(3)	&0.0000166153777(7)\\
50.0	&0.5000	&-0.000000029456062098(6)	&-0.000034663750907(8)	&-2.0000147017986(7)	&2.000034562489611(2)	&0.0000147073323(7)

    \\ \hline \hline
  \end{tabular*}}
\end{table}